%% file: arxiv.tex
\documentclass[11pt]{article}

\usepackage[a4paper,margin=2.4cm]{geometry}
\usepackage{amsmath,amssymb,bm}
\usepackage{booktabs,array,multirow}
\usepackage{graphicx}
\usepackage{xcolor}
\usepackage{microtype}
\usepackage[numbers,super,sort&compress]{natbib}

\usepackage{hyperref}
\usepackage{enumitem}
\usepackage[ruled,vlined]{algorithm2e}
\usepackage{verbatim}
\usepackage{soul}
\hypersetup{colorlinks=true, citecolor=orange, urlcolor=blue, linkcolor=magenta}

\newcommand{\BiWCM}{\mathrm{BiWCM}}
\newcommand{\PBiECM}{\mathrm{P\mbox{-}BiECM}}
\newcommand{\BiECM}{\mathrm{BiECM}}

\newcommand{\biwcm}{BiWCM\xspace}
\newcommand{\bipecm}{P-BiECM\xspace}
\newcommand{\biecm}{BiECM\xspace}
\newcommand{\abiecm}{qBiECM\xspace}
\newcommand{\bicm}{BiCM\xspace}

\newcommand{\github}{\href{https://github.com/lbuffa/wbnm}{\texttt{wbnm}}}

\newcommand{\E}{\mathbb{E}}
\newcommand{\Prob}{\mathbb{P}}
\newcommand{\ind}{\mathbb{I}}

\title{\bf Beyond node strengths:\\the topological content of weighted bipartite networks}

\author{Daniele Cirulli$^{1,2,+}$, Lorenzo Buffa$^{1,2,+}$, Giulio Cimini$^{2,1}$,\\
Dario Mazzilli$^{1}$, Aurelio Patelli$^{1}$, Fabio Saracco$^{1,3,4}$\\[0.5em]
\small $^1$``Enrico Fermi'' Research Center, 00184 Rome, Italy\\
\small $^2$Department of Physics and INFN, University of Rome Tor Vergata, 00133 Rome, Italy\\
\small $^3$Institute for Applied Computing ``Mauro Picone'', National Research Council, 00185 Rome, Italy\\
\small $^4$IMT School for Advanced Studies, 55100 Lucca, Italy\\
\small $^+$These authors contributed equally to the work}
\date{}

\usepackage{xspace}
\usepackage{subcaption}
\usepackage{adjustbox}
\usepackage{float}
\usepackage{tocloft}
\usepackage{placeins}

\begin{document}
\maketitle


\begin{abstract}
Weighted bipartite networks encode the topology and intensity of interactions between two disjoint sets of nodes. 
However, it is unclear whether the observed pattern of connections is already explained by node interaction volumes or contains additional structural information. 
Here we introduce a hierarchy of maximum-entropy models for continuous weighted bipartite networks that preserves node strengths alone, strengths and global density, or strengths and the complete degree sequence. Across economic, technological, ecological, social and lexical systems, we find that strength constraints alone are generally insufficient to reproduce empirical structures. 
Yet the missing topological information is often low-dimensional: 
a single constraint controlling network density frequently provides the most parsimonious description, 
whereas networks with stronger residual degree heterogeneity require node-specific constraints. 
Beyond downstream inference, we also show how the choice of topological resolution changes which interactions and projected relationships are identified as statistically significant. 
Our results indicate that topology is essential, but often compressible, in the statistical modeling of weighted bipartite networks, and provide a scalable computational framework for fitting, sampling, model selection and validation within a common workflow.
\end{abstract}

\section*{Introduction}
Bipartite weighted networks describe systems in which interactions occur between two distinct classes of entities and vary in intensity \citep{Borgatti1997,Latapy2008}. 
Countries export different volumes of products \citep{Hidalgo2009,Tacchella2012}, plants receive repeated visits from pollinators \citep{Rezende2007,Suweis2013,Gracia-Lazaro2018}, users interact with varying intensity over online communities \citep{Filho2020,Flamino2023}, and words occur with different frequencies across documents \citep{Dhillon2001,Bail2016,Gerlach2018,Simonetti2025}. In each case, the data contain two related but distinct forms of information: the topology specifies which pairs interact, whereas the weights quantify the intensity of the observed interactions.

A central problem in the analysis of these systems is determining whether topology carries information that is not already encoded by node interaction volumes -- namely, their strengths. 
On one hand, strength values alone do not specify how weights are divided among existing links: the same volume can be realized through many weak interactions or a few strong ones, producing radically different topologies. On the other hand, in the absence of additional structural mechanisms, nodes with larger (smaller) strengths naturally tend to connect to more (less) partners. 
Whether the observed pattern of zero and positive weights can be explained by weighted marginals is therefore a statistical question requiring an explicit statistical benchmark.

Maximum-entropy network models provide a principled way to construct such benchmarks \citep{Park2004,Garlaschelli2008,Squartini2011,Cimini2019}. They define the least biased probability distribution over networks that reproduces selected quantities of an empirical system. 
The choice of these quantities (the constraints of the entropy maximization procedure) determines which information is treated as explanatory; deviations from the resulting ensemble identify empirical structures that cannot be attributed to that information. 
Hence, a strength-only constrained model can test whether node activity is sufficient to explain both presence and weights of individual interactions. Adding constraints on the number of edges, globally or node by node, instead treats topology as an additional source of information.

The distinction between weighted and topological information is known to be consequential in monopartite networks. Strength-constrained models \citep{Serrano2006,Krivitsky2012,Sagarra2013,Squartini2011} often predict networks that are much denser than observed, because entropy maximization tends to spread node activity across all possible links, whereas jointly constraining degrees and strengths can greatly improve reconstruction \citep{Mastrandrea2014,Squartini2015,DiGangi2018}. Intermediate heuristic schemes further showed that strengths can be complemented by limited topological information, such as the link density of the network \citep{Cimini2015a,Cimini2015b} or an estimate of it \citep{Squartini2017b}. 
More recently, researchers have distinguished integrated from conditional mechanisms for generating links and assigning weights \citep{Gabrielli2019,Parisi2020,DiVece2023} and developed scalable algorithms for parameter inference \citep{Vallarano2021}. 

The corresponding problem is less developed for bipartite networks. While maximum-entropy methods have been studied extensively for binary bipartite data and their one-mode projections \citep{Saracco2015,Gualdi2016,Saracco2017,Cimini2022}, existing continuous-weight models are limited to strength-only formulations \citep{Bruno2023}, possibly adding the cost of establishing connections \citep{Buffa2025}. Consequently, it is unclear not only how topology can be constrained, but also whether this is needed to model empirical networks and at what resolution: is global density sufficient, or must the connectivity of every node be retained?

Here we answer this question by developing a unified nested hierarchy of canonical maximum-entropy models for continuous weighted bipartite data. The Bipartite Weighted Configuration Model (\(\BiWCM\)) constrains the strength sequences of the two sets of nodes and leaves topology unconstrained. The Partial Bipartite Enhanced Configuration Model (\(\PBiECM\)) additionally constrains the total number of links, introducing one parameter for global edge density. The Bipartite Enhanced Configuration Model (\(\BiECM\)) jointly constrains the strength and degree sequences of both sets, preserving node-specific connectivity.\footnote{A complementary microcanonical approach has recently been proposed to sample integer-weighted bipartite networks while preserving degree and strength sequences exactly \citep{Glaviano2026}.} This hierarchy allows us to determine the minimum topological information required for an adequate statistical description of a given empirical system.  

Across networks from five domains, we find that strengths alone are generally insufficient. Remarkably, however, topology does not always need to be specified node by node: a single density constraint often captures most of the information missing from the strength-only ensemble and is favored after penalizing model complexity. Local degree constraints become necessary when empirical connectivity remains heterogeneous after conditioning on strengths and density. 
Synthetic benchmarks with controlled density, heterogeneity and mesoscale organization reproduce this transition. These results establish topology as an essential, but often compressible, component of weighted bipartite network reconstruction. They also show that selecting an inadequate structural resolution can change which interactions and projected relations are considered statistically significant.

\section*{Results}

\subsection*{A hierarchy of hypotheses about topology}

Let \(W=(w_{i\alpha})\) be a non-negative weighted biadjacency matrix connecting $N_\top$ nodes \(\{i\}\) in layer \(\top\) to $N_\bot$ nodes \(\{\alpha\}\) in layer \(\bot\). Its binary support is \(b_{i\alpha}=\ind_{\mathbb{R}^+}(w_{i\alpha})\) ($\ind$ is the indicator function), hence the number of links is \(L=\sum_{i,\alpha}b_{i\alpha}\). Node strengths and degrees are
\begin{equation}
 s_i=\sum_\alpha w_{i\alpha},\qquad
 \sigma_\alpha=\sum_i w_{i\alpha},\qquad
 k_i=\sum_\alpha b_{i\alpha},\qquad
 h_\alpha=\sum_i b_{i\alpha}.
\end{equation}
The three maximum entropy models we consider retain the same weighted marginals but progressively enrich the description of topology (Table~\ref{tab:hierarchy}). They therefore encode competing hypotheses, rather than merely increasingly complicated approximations to the same object.

Full model derivations are given in Methods and in Section~\ref{sec:S2} of the Appendix. In a nutshell, \(\BiWCM\) associates an exponential weight distribution to every link, hence describes a fully connected system. 
The two enhanced models instead combine an exponential density for positive weights with an atom at zero: their link probability depends jointly on topological and weighted information. 

\begin{table}[h!]
\centering
\begin{tabular}{lcccc}
\toprule
Model &  \# links \(L\) & degrees \(\{k_i\},\{h_\alpha\}\) & \# parameters $q$\\
\midrule
\(\BiWCM\)   & - & - & $N_\top + N_\bot$\\
\(\PBiECM\)  & \checkmark & - & $N_\top + N_\bot + 1$\\
\(\BiECM\)   & implied & \checkmark & $2(N_\top + N_\bot)$\\
\bottomrule
\end{tabular}
\caption{\small Hierarchy of continuous weighted bipartite models. Check marks denote independently constrained quantities. All models preserve the strength sequences; they differ in whether topology is unconstrained, represented by global density, or specified through the two degree sequences.}
\label{tab:hierarchy}
\end{table}

\subsection*{Model selection}

Models are fitted on a given empirical network by maximum likelihood (see Methods and Section~\ref{sec:S2} of the Appendix for more details and the resolution correction used to compare likelihoods). 
Comparison among models $\{m\}$ is then performed using such maximized likelihood $\mathcal{L}_m$ as well as the Bayesian Information Criterion (BIC):
\begin{equation}\label{eq:bic}
  \mathrm{BIC}_m=q_m\log(N_\top N_\bot)-2\log\mathcal{L}_m,
\end{equation}
where \(q_m\) is the effective number of free parameters and $N_{\top}N_{\bot}$ is the number of dyads. 
Lower BIC values indicate a better balance between goodness of fit and model complexity. Further details can be found in Methods and Section~\ref{sec:S2} of the Appendix.

We investigated the performance of the various models upon a large corpus of economic, technological, ecological, social, and lexical systems (see Methods and Section~\ref{sec:S1} of the Appendix for further details on the empirical networks). For clarity, the economic, social, and technology datasets are repeated observations (for instance, the bipartite representation of international trade over different years), not independent systems.

\begin{figure}[t]
	\centering
	\includegraphics[width=\textwidth]{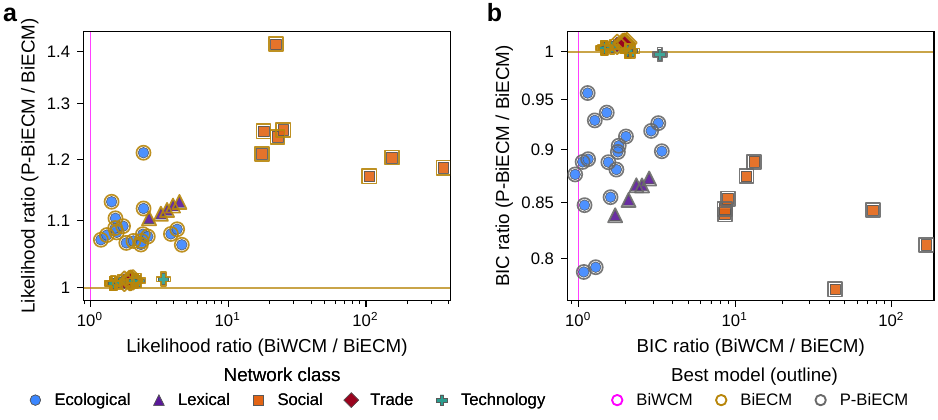}
	\caption{\small \textbf{Model comparison across empirical weighted bipartite networks.}
		Likelihood (a) and BIC (b) ratios of \(\BiWCM\) vs \(\BiECM\) (horizontal axis) and of \(\PBiECM\) vs \(\BiECM\) (vertical axis); \(\BiECM\) is used as the benchmark for the comparison, as it has the largest set of constraints and thus the highest likelihood, by definition.
		Each point represents one of the 46 empirical networks, coloured according to the class it belongs to; the outline marks the best model, i.e.\ the one with the highest likelihood (a) or the lowest BIC (b).
		Horizontal and vertical lines mark the values of 1, where models become equivalent.
		Because log-likelihoods are negative, the likelihood ratios are greater than 1, and the farther from 1, the worse the numerator model performs relative to \(\BiECM\). For BIC, ratios are greater than 1 when the model in the numerator performs worse than \(\BiECM\) and smaller when it performs better (because the lower the BIC, the more effective the model).
		While both \(\BiWCM\) and \(\PBiECM\) have lower likelihoods, BIC selects \(\PBiECM\) in 32 networks and \(\BiECM\) in 14, while \(\BiWCM\) is never selected.}
	\label{fig:model_selection}
\end{figure}

Figure~\ref{fig:model_selection} pictorially reports likelihood and BIC ratios among \(\BiWCM\), \(\PBiECM\) and \(\BiECM\) for our real data. We first assess that strength sequences alone cannot provide an adequate statistical description of the considered systems: \(\BiWCM\) is consistently disfavored relative to \(\BiECM\), even after accounting for its lower number of parameters. 
Once strengths-alone modeling is rejected, the relevant question is how much topology must be retained. The \(\PBiECM\) adds only one independent topological constraint, whereas \(\BiECM\) adds the heterogeneous degree sequences of both layers. As our hierarchy is genuinely nested (\(\PBiECM\) is a limit case of \(\BiECM\)), the raw likelihood necessarily favors the more constrained model, but the improvement is not always sufficient to offset its parameter cost.

In ecological, social and lexical networks, \(\PBiECM\) frequently provides the smallest BIC, meaning that the degree sequences information can be compressed, for model-selection purposes, into one density constraint. Economic and technological matrices instead more often favor \(\BiECM\), indicating that local connectivity contains information not captured by strengths and density alone.


The domain labels summarize the observed regimes but do not explain them. Several attempts were made to relate model preference and cross-model disagreement to elementary properties of the input network (see Section~\ref{sec:str_dr} of the Appendix), but no clear trend allowing a reliable a priori criterion was observed. 

\subsection*{Network reconstruction}
\begin{figure}[t!]
	\centering
	\includegraphics[width=0.6\textwidth]{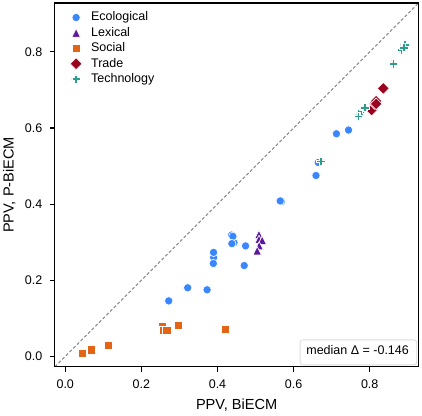}
	\caption{\small \textbf{Reconstruction of the observed links.} Comparison between \(\PBiECM\) and \(\BiECM\) in the reconstruction of the observed links of the 46 empirical networks, measured through the positive predictive value $\mathrm{PPV}=\langle\mathrm{TP}\rangle/(\langle\mathrm{TP}\rangle+\langle\mathrm{FP}\rangle)$, the expected fraction of predicted links that are observed (Methods). The dashed line is the identity; the median difference between \(\PBiECM\) and \(\BiECM\) is $-0.146$. The specificity, i.e.\ the fraction of empirical zeros predicted as zeros, is reported in Section~\ref{sec:reconstruction} of the Appendix.}
	\label{fig:zeros}
\end{figure}

We now test the model's ability to reproduce the empirical topology in terms of individual links. 
For each model, the link probability is jointly determined by the imposed constraints (see Methods).  
Figure~\ref{fig:zeros} shows that the Positive Predictive Value PPV for the existence of links obtained with \(\PBiECM\) is systematically lower than the one obtained with \(\BiECM\). 
The difference is appreciable for some ecological and social networks, while trade and technological systems tend to display higher PPV under both models. 
Thus, preserving the global density captures only part of the information contained in the empirical support: node-specific degree constraints improve the ability of the model to identify where interactions actually occur. 
Note that \(\BiWCM\) is not included in this comparison as it basically cannot predict non-existing links.

\begin{figure}[t!]
	\centering
	\includegraphics[width=\textwidth]{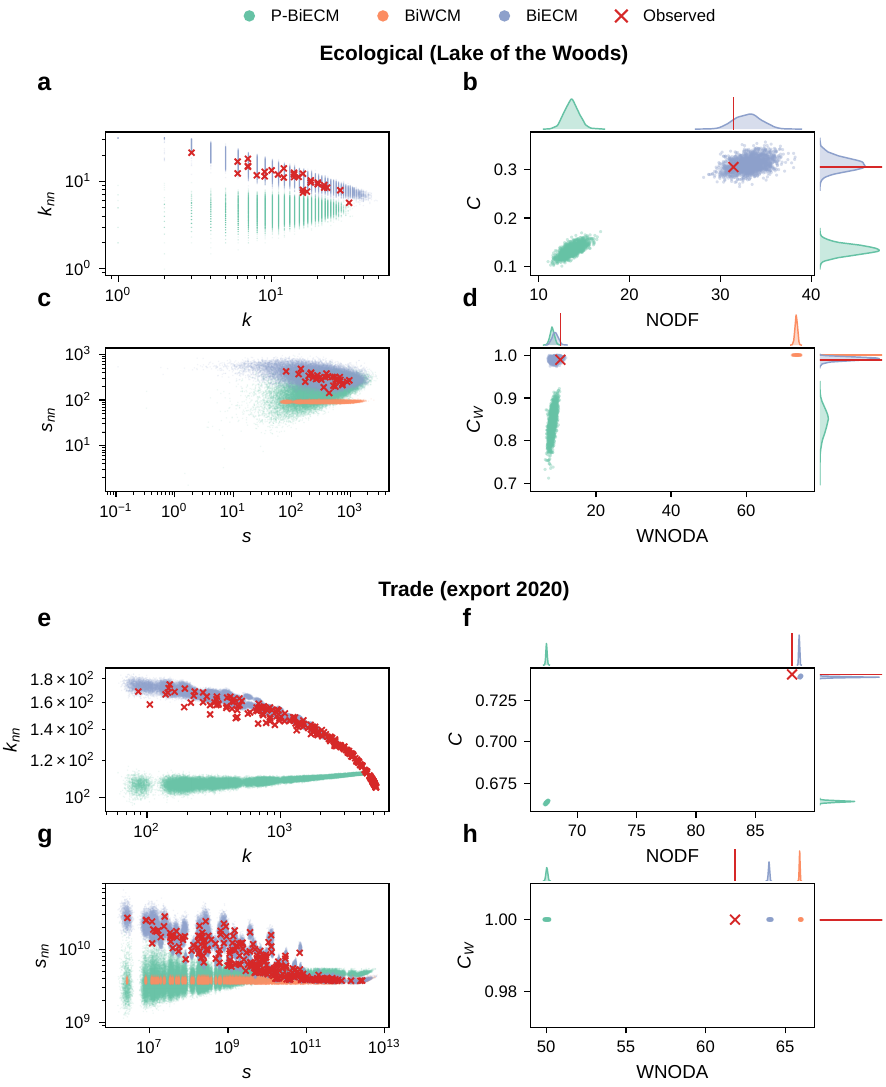}
	\caption{\small \textbf{Reconstruction of unconstrained structural properties} for an ecological network (Lake of the Woods, \textbf{a--d}) and a trade network (export 2020, \textbf{e--h}).
		\textbf{a},\textbf{e},~Mesoscopic reconstruction of the average nearest-neighbour degree $k_{nn}$ as a function of the node degree $k$, and \textbf{c},\textbf{g},~of the average nearest-neighbour strength $s_{nn}$ as a function of the node strength $s$ (one point per node and sample).
		\textbf{b},\textbf{f},~Macroscopic reconstruction of the binary structure, clustering $C$ against NODF, and \textbf{d},\textbf{h},~of the weighted structure, weighted clustering $C_W$ against WNODA, with the marginal distributions of the samples.
		Red crosses: observed values; $2\times10^3$ samples per model (definitions in Methods). The \(\BiWCM\) samples are complete graphs, so their binary observables are not informative and are not shown in \textbf{a}, \textbf{b}, \textbf{e} and \textbf{f}. In the dense trade network the weighted clustering saturates ($C_W\simeq1$) in the data and in all ensembles, so this observable does not discriminate between the models (\textbf{h}).}
	\label{fig:reconstruction}
\end{figure}

We next test models against properties not imposed as constraints, including nearest-neighbour degree and strength, clustering and nestedness (definitions in Methods). In the following, we present results for a representative of our set of ecological networks (Figure \ref{fig:reconstruction}, panels a-d), and a representative of our set of trade networks (Figure \ref{fig:reconstruction}, panels e-h): more results can be found in Section~\ref{sec:S6} of the Appendix.

As shown in all panels of Figure \ref{fig:reconstruction}, \biecm generally provides the most accurate reconstruction of purely topological observables, as expected given that it accounts for the full degree sequences. For observables combining topology and weight, however, the relative performance of the models is less consistent across systems, as illustrated by comparing panels c and d with panels g and h. In the country-product matrix, for instance, the observed nearest-neighbour strength ($s_\text{nn}$) patterns lie close to the \biecm expectations (panel g), suggesting that the model captures the local connectivity heterogeneity of the system. A similar trend emerges for ecological networks, although the expected distribution of $s_\text{nn}$ is broader (panel c).

More pronounced differences emerge for other weighted observables. In ecological networks, both \biecm and \bipecm reproduce weighted nestedness (WNODA) reasonably well, whereas only \biecm effectively reconstructs weighted clustering ($C_W$; panel d). In international trade, the pattern is almost reversed (panel h): all models effectively reconstruct $C_W$, but none reproduces WNODA. Moreover, the WNODA results are statistically significant for all models, with the corresponding deviations falling on the right-hand side of the reference distributions for \biecm and on the left-hand side for \bipecm.

These results highlight system-specific structural features that remain unexplained by the selected low-order constraints. In international trade, reproducing the degree and strength sequences is insufficient to capture weighted nestedness, pointing to the need for additional information about the hierarchical organisation of export baskets. By contrast, the higher-order patterns measured by $C_W$ are already well captured by the models. In ecological networks, the picture is different: weighted nestedness is reproduced comparably well by both \biecm and \bipecm, whereas the constraints encoded in \bipecm are insufficient to reproduce weighted clustering, which is effectively captured only by \biecm.

This comparison complements, but should not be conflated with, information-theoretic model selection. A model may improve a particular observable while being disfavoured after accounting for its additional parameters. Conversely, selecting \(\PBiECM\) by BIC means that its loss of likelihood is not sufficient to justify the full degree sequences; it does not imply that \(\PBiECM\) reproduces every topological statistic as accurately as \(\BiECM\).


\subsection*{Statistical validation}
The effect of including the topological information in the model may have deep implications. 
Maximum-entropy ensembles are frequently used not only to statistically model a network but also to identify patterns that are unlikely under a specified null hypothesis: that the imposed constraints explain everything of the empirical system \citep{Cimini2019}. When this null is rejected, it means that the detected pattern carries genuinely additional information beyond that used to define the model. 
Therefore, the choice of the null model and the ingredients it should include are fundamental for the conclusions to be deduced from the statistical validation. 

\subsubsection*{Weights validation} 
\begin{figure}[!t]
	\centering
	\includegraphics[width=\textwidth]{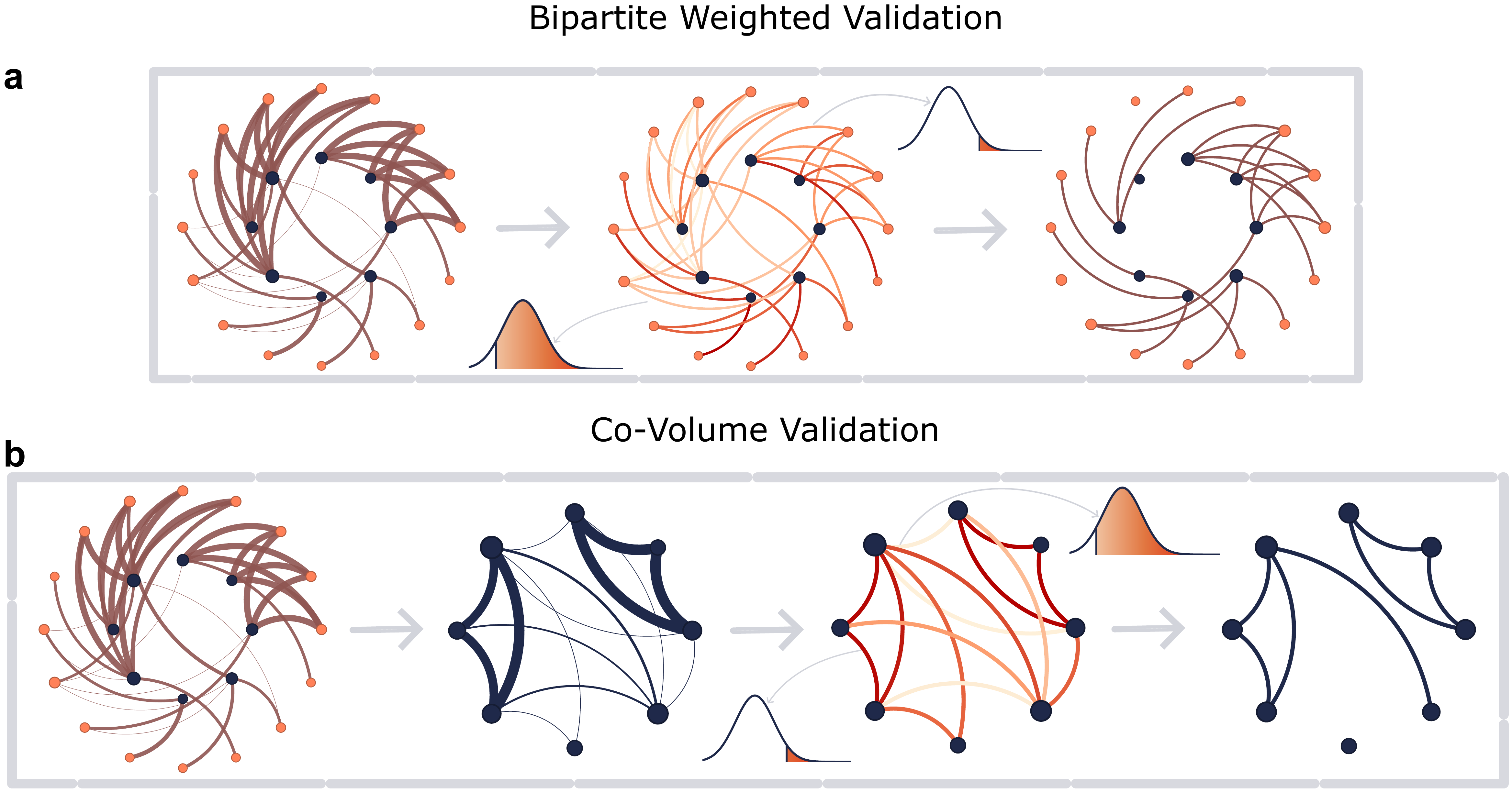}
	\caption{\small \textbf{Statistical validation of bipartite weights and co-volumes.}
		\textbf{a},~Each observed weight is compared with its distribution under the null model (analytical upper-tail p-value), and the links with a significant p-value are retained.
		\textbf{b},~The one-mode projection gives the co-volume of each pair of nodes; its null distribution is obtained by sampling the bipartite ensemble, and the pairs with a significant co-volume form the validated projection.}
	\label{fig:validation_schema}
\end{figure}

\begin{figure}[!t]
	\centering
	\includegraphics[width=1.00\textwidth]{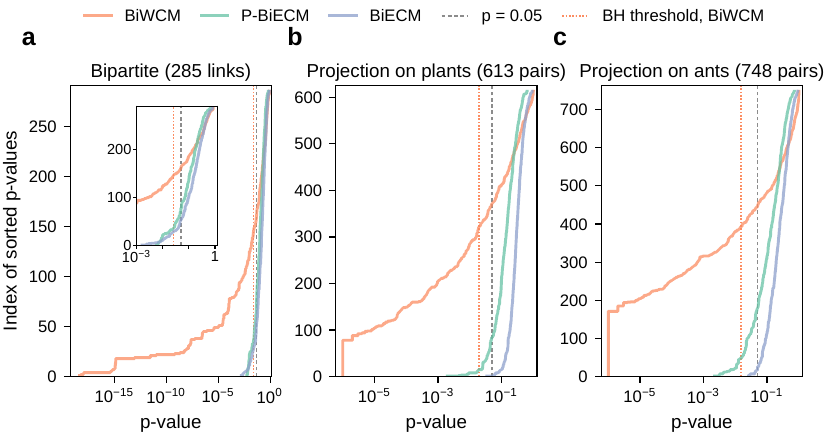}
	\caption{\small \textbf{Validation of weights and co-volumes in the plants--ants network} (41 plants, 51 ant species).
		Sorted p-values under the \(\BiWCM\), the \(\PBiECM\) and the \(\BiECM\) for \textbf{a},~the 285 observed links (inset: zoom on p-values between $10^{-3}$ and 1), and for the co-volumes of the projections on \textbf{b},~plants and \textbf{c},~ants (pairs with non-zero co-volume; $10^6$ samples per model). Dashed line: p-value $=0.05$; dotted line: Benjamini--Hochberg threshold (FDR 5\%) of the \(\BiWCM\).
		With the Benjamini--Hochberg procedure both the \(\PBiECM\) and the \(\BiECM\) validate no link. The dashed line is shown only as a reference.}
	\label{fig:plants_ants}
\end{figure}

We start with the statistical validation of link weights. The procedure is pictorially represented in panel a of Figure~\ref{fig:validation_schema}. For each observed positive weight, we computed its p-value (upper-tail probability) under each ensemble. The p-value calculation is analytical in the Lagrangian multipliers and, thus, it does not need any sampling (see Methods). P-values are then validated, controlling the false discovery rate across dyads \citep{fdrBenjamini1995} (again, see Methods for further details). Panel a of Figure \ref{fig:plants_ants} shows the p-value profile of the various models for an ecological mutualistic network between plants and ants: the same empirical matrix produces different p-value distributions and validated interaction sets under the three null hypotheses. Qualitatively, the same conclusions can be drawn for other empirical networks in our corpus.

The different profiles highlight how the resolution of topological constraints changes link-level statistical validation.
The strength-only \biwcm generally labels more interactions as anomalous (i.e. smaller p-values) because it does not reproduce empirical sparsity. Adding global density makes the benchmark more selective (i.e. greater p-values). Furthermore, adding node-specific degrees produces a further reduction where local connectivity is heterogeneous, as the profile of the p-value curve is lower for \biecm than \bipecm. Across the empirical networks, \bipecm and \biecm agree more closely with one another than either does with \biwcm, but their remaining disagreement is systematic rather than negligible. 


\subsubsection*{Co-volumes validation}
We finally apply statistical validation to empirical co-volumes resulting from one-mode projections of the bipartite systems. The choice to validate projected relationships through a null model defined on the bipartite system, rather than one estimated directly on the projected co-volume matrix, follows from two considerations. 

First, the constraints that define the null hypothesis are quantities with a direct empirical meaning only at the bipartite level: they describe the activity of the original entities before projection, whereas co-volumes are derived quantities with no corresponding directly-observed constraint. A null model fitted directly on the projection would therefore include information from a derived object (i.e. the projection), instead of focusing on principled proxies for any generative mechanism of the underlying system. 

Second, such a model offers no guarantee of preserving the topological and weighted structure of the bipartite network the projection originates from: because projection is a many-to-one, non-linear transformation of the biadjacency matrix, bipartite networks with markedly different structure (such as different degree heterogeneity, or a different resolution of topological information, as discussed above) can yield projections with similar co-volume distributions. A null model calibrated only on the projection would be blind to this upstream information, so the resulting validated relationships need not reflect the mechanisms that actually generate the observed pattern in the real bipartite system \citep{Saracco2017,Cimini2022}. Deriving the null distribution of co-volumes from the bipartite ensemble instead allows us to ask a sharper question: are the observed projected relationships expected once the bipartite generative mechanism has been properly accounted for?

The procedure is pictorially represented in panel b of Figure~\ref{fig:validation_schema}. Unlike the validation of the bipartite weights, due to the intricate dependencies, we could not find a closed-form formula for the calculation of the p-values of co-volumes. Therefore, they were calculated by sampling from the various ensembles. 

As can be expected, the null distribution of the co-volume matrix and the resulting validated projection are strongly model-dependent. In panels b and c of Figure \ref{fig:plants_ants}, the p-value profiles for the various models are presented for the projection on the plants and the ants layer, respectively ($10^6$ samples per model). Similarly to what was observed for the weight validation, ignoring the topology makes many observations surprising for \(\BiWCM\), as in the case of the weight validation above. Otherwise stated, the co-volume distribution predicted by the models including the topological information (i.e. \(\PBiECM\) and \(\BiECM\)) is narrower and closer to the observed values. Differently for the weighted validation, nonetheless, the profiles of \bipecm and \biecm are well resolved, stating that the difference in the resolution of the topology is relevant when studying the co-volume distribution. Similarities and differences in the co-volume validation can be found in Sections~\ref{sec:S6},~\ref{sec:io},~\ref{sec:synth} of the Appendix.



\section*{Discussion}

Across the systems studied here, strength-constrained ensembles fail to account for empirical support, showing that the allocation of positive and zero interactions carries additional information. The important distinction, however, is not simply between models with and without topology. It is between the resolutions at which topology becomes informative.

The hierarchy developed here turns this distinction into a model-selection problem. The \(\PBiECM\) asks whether one global sparsity constraint is sufficient beyond the two strength sequences; the \(\BiECM\) asks whether the connectivity of each node must additionally be retained. In many empirical systems, global density is the most economical explanation of the departure from a strength-only baseline. In others, broad or poorly predicted degrees make local constraints indispensable for an adequate description. Thus, accounting for topology does not necessarily require node-specific constraints.

This conclusion refines earlier results for monopartite weighted networks, where degree information was shown to improve reconstruction beyond strengths \citep{Mastrandrea2014}, and density-assisted procedures were shown to recover much of an unknown topology from limited information \citep{Cimini2015a,Squartini2017a}. Our contribution is to formulate and compare these hypotheses in a common canonical framework for continuous weighted bipartite data. It complements 
the recent microcanonical algorithm that preserves bipartite degrees and strengths exactly \citep{Glaviano2026}. Canonical and microcanonical approaches answer related but distinct questions: the former represents uncertainty around constraints satisfied on average, whereas the latter randomizes within an exactly constrained state space - the two approaches are not equivalent in the presence of local constraints \citep{Squartini2015a,Garlaschelli2016}.

Our results also expose a practical consequence of model choice. Validation of link weights and co-volumes is conditional on the null ensemble, even when applied to the same empirical matrix. A model that does not reproduce sparsity tends to overstate the unexpectedness of observed quantities. Conversely, adding local degree constraints can remove relationships whose apparent significance follows from heterogeneous connectivity. Validated networks should therefore be reported together with, and justified by, an explicit selection of their underlying null model.

Several limitations remain. First, the empirical corpus contains different temporal realizations of the same dataset and uneven numbers of networks across domains. Domain-level conclusions should therefore be interpreted at the level of independent systems rather than raw matrix counts. 
Second, the present hierarchy stops at degree constraints. Mesoscale organization, spatial costs, or covariates may contain additional information and can be incorporated as further constraints when justified by the application.

Overall, the proposed model hierarchy provides a quantitative notion of topological compressibility. If one density parameter leads to a performance comparable to that obtained using two complete degree sequences after accounting for complexity, most low-order topological information is reducible to global sparsity. If it does not, local connectivity is an unavoidable component of the statistical description. 

\section*{Methods}

\subsection*{Data}

We analysed weighted bipartite matrices from economic, technological, ecological, social and lexical systems. Economic networks connecting countries to exported products are derived from UN-COMTRADE~\citep{JRC136943}. 
Technological networks connect countries with technological CPC classes from patent inventors' geolocations obtained from the PATSTAT database~\citep{PATELLI20171229}. 
Ecological networks encode observed mutualistic or trophic interactions, obtained from public interaction databases, including IWDB and ICON \citep{IWDB,ICON}. 
Social networks connect users to online communities or information sources, built from Reddit and related online-interaction datasets \citep{baumgartner2020pushshift,cirulli2025polarizationechochambersreddits}. 
Lexical networks connect documents to processed word stems for UK companies \citep{Carungu2026}. 
Basic statistics, data sources, and all preprocessing operations are reported in Section~\ref{sec:S1} of the Appendix.

\subsection*{Maximum-entropy ensembles}

\subsubsection*{General framework}
Given a real network $G^*$ (in the following, all quantities denoted by an asterisk are measured on the empirical network), we first define $\mathcal{G}$, the set of all possible graphs having the same number of nodes as in the empirical system, from the empty graph to the fully connected one. We equip each representative of this set with a probability obtained through a constrained maximization of the Shannon entropy, defined as
\begin{equation}\label{eq:Shannon_entropy}
    S=-\sum_{G\in\mathcal{G}} P(G)\ln P(G).
\end{equation}

Let us call $\vec{C}$ the vector of the constraints. The result of the constraint maximization of the Shannon entropy in Eq.\eqref{eq:Shannon_entropy} returns a probability for each representative of $\mathcal{G}$ that depends on the value of $\vec{C}$ measured on $G$:
\begin{equation*}
	P(G|\vec{\theta})=
	\frac{\exp[-\mathcal{H}(G,\vec{\theta})]}{Z(\vec{\theta})},
	\qquad
	\mathcal{H}(G,\vec{\theta})=\vec{\theta}\cdot\vec{C}(G)= \sum_m \theta_m C_m(G),
\end{equation*}

where $Z$ is the partition function and $\vec{\theta}$ is the vector of Lagrange multipliers. 
Parameters are estimated by imposing that the average value over the ensemble of the constraints is equal to the value measured on the empirical network:
\begin{equation*}
	\langle \vec{C}\rangle=\vec{C}({G}^*).
\end{equation*}
Remarkably, such a condition is equivalent to the maximization of the likelihood; such an equivalence is valid only for maximum entropy models~\citep{Garlaschelli2008}. 

As the constraints considered in the present paper are linear either in the binary biadjacency matrix $\mathbf{B}$ or in the weighted biadjacency matrix $\mathbf{W}$, the resulting probability distributions per graph factorize over probability distributions per node pairs~\citep{Park2004,Garlaschelli2008}.

\subsubsection*{Bipartite Weighted Configuration Model.} 
The Bipartite Weighted Configuration Model (\(\BiWCM\)) constrains the strength sequences for each layer, i.e. \(\{s_i\}\) and \(\{\sigma_\alpha\}\). With corresponding multipliers \(\{\Theta_i\}\) and \(\{H_\alpha\}\), the weight of link $(i,\alpha)$ follows the distribution
\begin{equation}\label{eq:q_biwcm}
q^{\BiWCM}_{i\alpha}(w)=\lambda_{i\alpha}e^{-\lambda_{i\alpha}w},
\qquad \lambda_{i\alpha}=\Theta_i+H_\alpha,
\qquad w\geq0
\end{equation}
hence \(\langle w_{i\alpha}\rangle=1/\lambda_{i\alpha}\). As the distribution is continuous, it has no probability mass at zero; therefore, the \(\BiWCM\) controls weighted activity, but not the topology of the network.

\subsubsection*{Partial Bipartite Enhanced Configuration Model.}
\(\PBiECM\) constrains \(\{s_i\}\), \(\{\sigma_\alpha\}\), and \(L\). Let \(\delta\) be the multiplier associated with \(L\). The dyadic law is a hurdle distribution,
\begin{equation}\label{eq:q_p_biecm}
q^{\PBiECM}_{i\alpha}(w)=\left\{
\begin{array}{ll}
    1-p_{i\alpha} &  w=0\\
    p_{i\alpha}\lambda_{i\alpha}e^{-\lambda_{i\alpha}w} & w>0
\end{array}
\right.
\end{equation}
where $p_{i\alpha}$ is the probability of link existence,
\begin{equation}
p_{i\alpha}=\frac{e^{-\delta}}{\lambda_{i\alpha}+e^{-\delta}},
\end{equation}
and $\lambda_{i\alpha}=\Theta_i+H_\alpha$ (while the relation between $\lambda_{i\alpha}$, $\Theta_i$ and $H_\alpha$ is analogous to the one in Eq.\eqref{eq:q_biwcm}, the numerical values of the Lagrangian multipliers are different for the two models).
Although topology has one independent multiplier, connection probabilities vary across dyads through the strength multipliers entering \(\lambda_{i\alpha}\).

\subsubsection*{Bipartite Enhanced Configuration Model.}
\(\BiECM\) constrains \(\{s_i\}\), \(\{\sigma_\alpha\}\), \(\{k_i\}\), and \(\{h_\alpha\}\). Denoting by \(\{\theta_i\}\) and \(\{\eta_\alpha\}\) the multipliers of the latter two constraints, the functional form of the weight distribution is analogous to the one in Eq.~\eqref{eq:q_p_biecm}, i.e. 
\begin{equation}
q^{\BiECM}_{i\alpha}(w)=\left\{
\begin{array}{ll}
    1-p_{i\alpha} &  w=0\\
    p_{i\alpha}\lambda_{i\alpha}e^{-\lambda_{i\alpha}w} & w>0
\end{array}
\right.,
\end{equation}
but now the probability of link existence depends also on the multipliers for the degree sequences:
\begin{equation}\label{eq:p_biecm}
p_{i\alpha}=\frac{e^{-(\theta_i+\eta_\alpha)}}{\lambda_{i\alpha}+e^{-(\theta_i+\eta_\alpha)}}.
\end{equation}
Due to the functional form of Eq.~\eqref{eq:p_biecm}, the topology and positive weights are inferred jointly because \(p_{i\alpha}\) depends on both topological and strength multipliers. 

In fact, solving the parameters of the \biecm can be computationally demanding. Therefore, in our toolkit, we provide the quasi-Bipartite Enhanced Configuration Model (\abiecm), i.e. a method inspired by the \biecm, but introducing a simplification in the derivation that dramatically reduces the computation time of the parameters. More details can be found in Section~\ref{sec:qDECM} of the Appendix.

\subsubsection*{Weights validation} 
 
For each observed positive bipartite weight $w_{i\alpha}^*$, we compute a right-tail p-value under the selected null ensemble. For the \biwcm this tail probability is 
\begin{equation}
p^{\mathrm{link}}_{i\alpha}=\exp(-\lambda_{i\alpha}w_{i\alpha}^*).\label{eq:plink_biwcm}    
\end{equation} 
For the topology-aware weighted models, i.e. \biecm and \bipecm, the probability also includes the chance that the edge exists,
\begin{equation}
p^{\mathrm{link}}_{i\alpha}=p_{i\alpha}\exp(-\lambda_{i\alpha}w_{i\alpha}^*).\label{eq:plink_hurdle}
\end{equation}
This distinction is central when testing sparse weighted matrices, because topology-aware models can assign probability to empirical zeros whereas the \biwcm cannot. 
Because we have a p-value for each pair $(i,\alpha)$, 
the p-values are corrected for multiple hypothesis testing. The package implements both the Bonferroni
correction and False Discovery Rate control; all the results use the latter, with the
Benjamini--Hochberg procedure at significance level $\alpha_{\mathrm{sig}}=0.05$ (see the
``Multiple-hypothesis testing'' paragraph below).

\subsubsection*{Co-volumes validation} 
We used the same framework to validate one-mode projections. For each pair of nodes in the projected layer, we compute an empirical weighted co-volume $V_{ij}^*$ from their shared neighbours in the opposite layer. 
Since there is not a simple closed-form formula for the distribution of co-volumes, the null distribution of $V_{ij}$ is obtained by sampling from the fitted bipartite ensemble. The projection p-value is
\begin{equation}
p^{\mathrm{proj}}_{ij}=\Prob\left(V_{ij}^{\mathrm{null}}\ge V_{ij}^*\right).\label{eq:pproj}
\end{equation}

It is estimated as the fraction of the $S$ sampled networks whose co-volume is at least the observed one,
$\widehat p_{ij}=S^{-1}\sum_{s=1}^{S}\ind\big(V^{(s)}_{ij}\ge V^*_{ij}\big)$. As for the links, the p-values
are corrected for multiple hypothesis testing with the Benjamini--Hochberg procedure at
$\alpha_{\mathrm{sig}}=0.05$. When the procedure validates no pair for at least one model, which happens for
all the ecological networks, all the models of that network are compared with the same fixed threshold,
$\widehat p_{ij}<0.05$ (Section~\ref{sec:reconstruction} of the Appendix).

\subsection*{Finite observation resolution and comparable likelihoods}
In Section~\ref{sec:S2} of the Appendix, the interested reader can find the explicit form of the log-likelihood densities $\ell_\text{model}$, calculated for each of our models. Here we remark a non-trivial detail necessary for the understanding of the model comparison. 

When calculating BIC, see Eq.\eqref{eq:bic}, the log-likelihood appears. In our models, we have probability densities with respect to a mixed continuous measure instead and cannot be compared directly with probabilities assigned to recorded observations. Thus, comparing all three models on a common footing requires converting the continuous density into an actual probability. 

In fact, empirical weights are recorded with finite resolution \(\Delta w\). We therefore compare models using probabilities integrated over the observation bins. 
For instance, for \biwcm ,
\begin{equation}
\Prob^\text{\biwcm}(w_{i\alpha}=w)\simeq q_{i\alpha}^\text{\biwcm}(w)\frac{1-e^{-\lambda_{i\alpha}\Delta w}}{\lambda_{i\alpha}}.
\label{eqSI:binprob}
\end{equation}
Thus, the discretized \(\BiWCM\) is a geometric distribution over the recorded bins and assigns a non-zero probability to an observed zero through the interval \([0,\Delta w)\). For any of our topology-aware models, applying the same trick gives
\begin{equation}
\Prob(w_{i\alpha}=0)=1-p_{i\alpha}e^{-\lambda_{i\alpha}\Delta w},
\qquad
\Prob(w_{i\alpha}=w>0)=\dfrac{1-e^{-\lambda_{i\alpha}\Delta w}}{\lambda_{i\alpha}}p_{i\alpha}\lambda_{i\alpha}e^{-\lambda_{i\alpha}w}
\label{eq:hurdlebinprob}
\end{equation}

These corrections add a sum of terms to $\ell$, over every dyad for the purely continuous \biwcm, and only over the dyads where a link is actually observed ($b_{i\alpha}=1$) for the topology-aware \bipecm and \biecm:
\begin{align}
\ln\mathcal{L}_{\text{\biwcm}}&=\ell_{\text{\biwcm}}+\sum_{i,\alpha}\ln\left(\frac{1-e^{-\lambda_{i\alpha}\Delta w}}{\lambda_{i\alpha}}\right), \label{eqSI:ll_biwcm_corrected}\\
\ln\mathcal{L}_{\text{\bipecm}}&=\ell_{\text{\bipecm}}+\!\!\sum_{(i,\alpha)|\,b_{i\alpha}=1}\!\!\ln\left(\frac{1-e^{-\lambda_{i\alpha}\Delta w}}{\lambda_{i\alpha}}\right)+\!\!\sum_{(i,\alpha)|\,b_{i\alpha}=0}\!\!\ln\left(\frac{1-p_{i\alpha}e^{-\lambda_{i\alpha}\Delta w}}{1-p_{i\alpha}}\right), \label{eqSI:ll_bipecm_corrected}\\
\ln\mathcal{L}_{\text{\biecm}}&=\ell_{\text{\biecm}}+\!\!\sum_{(i,\alpha)|\,b_{i\alpha}=1}\!\!\ln\left(\frac{1-e^{-\lambda_{i\alpha}\Delta w}}{\lambda_{i\alpha}}\right)+\!\!\sum_{(i,\alpha)|\,b_{i\alpha}=0}\!\!\ln\left(\frac{1-p_{i\alpha}e^{-\lambda_{i\alpha}\Delta w}}{1-p_{i\alpha}}\right). \label{eq:ll_biecm_corrected}
\end{align}

The log-likelihoods and BIC values reported in the main text are computed with this exact, integrated correction throughout, so that likelihoods are always compared on the same footing across the hierarchy, including on the exact zeros where the shortcut would otherwise fail silently.

\subsection*{Multiple-hypothesis testing}


When handling a great number of p-values, multiple-testing corrections are required to control false
positives. Let $p_1,\dots,p_D$ be the p-values of the $D$ tested hypotheses ($p^{\mathrm{link}}_{i\alpha}$ for
the links, $\widehat p_{ij}$ for the projections) and $\alpha_{\mathrm{sig}}$ the significance level.
Under Bonferroni correction \citep{bonferroni1936teoria}, hypothesis $r$ is rejected if
\begin{equation}\label{eq:bonf}
p_r \le \frac{\alpha_{\mathrm{sig}}}{D}.
\end{equation}
Under False Discovery Rate criterion \citep{fdrBenjamini1995}, once the p-values are sorted in increasing
order, $p_{(1)}\le p_{(2)}\le\dots\le p_{(D)}$, one identifies the greatest rank $r^*$ satisfying
\begin{equation}\label{eq:bh}
p_{(r^*)}\le \frac{r^*\alpha_{\mathrm{sig}}}{D}
\end{equation}
and rejects all hypotheses with p-values smaller than or equal to
$t_{\mathrm{BH}}=r^*\alpha_{\mathrm{sig}}/D$. 

\subsection*{Topology and structural reconstruction}

\paragraph{Positive Predictive Value} For the topology-aware models the connection probabilities $p_{i\alpha}$ give the expected confusion matrix with respect to the observed support $b^*_{i\alpha}$: $\langle\mathrm{TP}\rangle=\sum_{b^*_{i\alpha}=1}p_{i\alpha}$ and $\langle\mathrm{FP}\rangle=\sum_{b^*_{i\alpha}=0}p_{i\alpha}$. The positive predictive value $\mathrm{PPV}=\langle\mathrm{TP}\rangle/(\langle\mathrm{TP}\rangle+\langle\mathrm{FP}\rangle)$ is the expected fraction of predicted links that are observed; it is an analytic expectation.

\paragraph{Assortativity measures} The average nearest-neighbour degree and strength of node $i$ are $k^{nn}_i=k_i^{-1}\sum_\alpha b_{i\alpha}h_\alpha$ and $s^{nn}_i=k_i^{-1}\sum_\alpha b_{i\alpha}\sigma_\alpha$, with $h_\alpha$ and $\sigma_\alpha$ the degree and strength of node $\alpha$ of the other layer. Their ensemble mean $\langle x_i\rangle$ and standard deviation $\sigma_{x_i}$ are Monte Carlo estimates from networks sampled from each fitted model, and the agreement with the data is summarised by $\mathrm{RMSE}_z=\big[N^{-1}\sum_i z_i^2\big]^{1/2}$, with $z_i=(x^*_i-\langle x_i\rangle)/\sigma_{x_i}$, which is close to one when the observable is reproduced within the model's own fluctuations.

\paragraph{Nestedness} Nestedness is measured by NODF \citep{AlmeidaNeto2008}, the average over pairs of nodes ordered by decreasing degree of the overlap $|\Gamma_i\cap\Gamma_j|/k_j$ between their neighbourhoods, and by its weighted counterpart WNODA \citep{Pinheiro2019}, in which nodes are ordered by strength and the overlap counts the entries of $j$ that are smaller than those of $i$. 

\paragraph{Clustering} Binary clustering is the Robins--Alexander coefficient $C=4C_4/L_3$, the fraction of paths of length three closed into 4-cycles, computed with NetworkX \citep{Hagberg2008}; weighted clustering is the two-mode coefficient of Opsahl \citep{Opsahl2013}, the weighted fraction of closed 4-paths, with each path weighted by the geometric mean of its weights. 

For each of these observables, the observed value is compared with its sampled distribution through the z-score and the percentile of the observed value. Exact definitions, numbers of samples, and results are given in Section~\ref{sec:reconstruction} of the Appendix.

\subsection*{Pseudocode and numerical implementation}
The complete workflow, summarized in Algorithm~\ref{alg:wbnm}, is
implemented in an open-source Python package built on PyTorch, exposing
a common interface for the three null models:
fitting, sampling, and validation share the same routines across BiWCM,
P-BiECM, and BiECM. Because the empirical analysis above
requires fitting several competing models on each of many networks and
drawing large sampling ensembles ($S$ from $10^3$ to $10^6$ networks per model, see Section~\ref{sec:reconstruction} of the Appendix) for both reconstruction
and projection validation, computational efficiency is a design
requirement, not an afterthought, in the spirit of recent scalable
likelihood-maximization solvers for constrained network
ensembles~\citep{Vallarano2021}. All solvers are fully vectorized over
dyads, can be chunked row-wise to bound memory use on large matrices,
and are moved to GPU when explicitly requested via a device argument.

Parameter fitting solves the moment-matching equations underlying
$\langle \vec{C}\rangle=\vec{C}({G}^*)$.
The default solver is a fixed-point iteration with
Polyak-style momentum (step size $\gamma=1$, momentum $\beta=0.9$ by
default), used whenever degree and strength constraints are solved
jointly (P-BiECM, BiECM);
momentum can optionally be annealed toward $\beta=0.999$ over a
warm-up window to stabilize early iterations without slowing
late-stage convergence. The single-constraint BiWCM converges without
momentum.
A diagonal quasi-Newton update, with a model-specific step
size tuned for stability, and an exact bisection scheme ($60$ steps in
log-space, $\sim10^{-18}$ precision) are provided as alternatives, the
latter offering guaranteed monotonic convergence when the fixed-point
iteration is unstable.
Convergence is assessed on the maximum relative
error between empirical and expected constraints: in all the analyses of this work a solution is accepted when this error is below $\varepsilon=10^{-3}$ (the package default, $\varepsilon=10^{-8}$ with at most $10^4$ iterations, is stricter and can be set by the user). Further details can be found in Section~\ref{sec:numerical} of the Appendix.

\begin{algorithm}[t]
\small
\caption{WBNM fitting--sampling--validation pipeline}
\label{alg:wbnm}
\SetAlgoLined
\DontPrintSemicolon
\KwIn{$\mathbf{W}$; model $\mathcal{M}\in\{\text{BiWCM, P-BiECM, BiECM,
qBiECM}\}$ (Table~\ref{tab:hierarchy}); tolerance $\varepsilon$; number of samples
$S$; significance level $\alpha_{\mathrm{sig}}$}
\KwOut{Fitted multipliers, BIC, validated bipartite links, validated
projection}

\tcp{1. Multiplier resolution}
$C^* \leftarrow$ empirical constraints of $\mathbf{W}$ (Eq.~1)\;
initialize $\{\Theta,H,\theta,\eta\}^{(0)}$; $v\leftarrow0$;
$t\leftarrow0$\;
\Repeat{$\max|\Delta^{(t-1)}|<\varepsilon$ \textnormal{or} $t=t_{\max}$}{
  $\Delta^{(t)} \leftarrow \langle C\rangle^{(t)}/C^* - 1$ \tcp*{vectorized, chunked dyads (CPU/GPU)}
  $v \leftarrow \beta_t v + (1-\beta_t)\,\Delta^{(t)}$\;
  $\{\Theta,H,\theta,\eta\}^{(t+1)} \leftarrow
  \{\Theta,H,\theta,\eta\}^{(t)} \odot (1+\gamma v)$ \tcp*{or quasi-Newton / bisection}
  $t \leftarrow t+1$\;
}

\tcp{2. Model selection and link validation}
$\mathcal{L}, \mathrm{BIC} \leftarrow \textsc{Likelihood}(\mathbf{W},\{\Theta,H,\theta,\eta\})$ \tcp*{Eq.~\eqref{eq:bic}}
$p^{\mathrm{link}}_{i\alpha} \leftarrow \textsc{PValue}(w^*_{i\alpha},\{\Theta,H,\theta,\eta\})$ for every observed link \tcp*{Eqs.~\eqref{eq:plink_biwcm}, \eqref{eq:plink_hurdle}}

\tcp{3. Sampling: reconstruction and projection validation}
\For{$s=1$ \KwTo $S$}{$\mathbf{W}^{(s)} \leftarrow
\textsc{Sample}(\mathcal{M}, \{\Theta,H,\theta,\eta\})$; $V^{(s)}\leftarrow V(\mathbf{W}^{(s)})$\;}
$\widehat p_{ij} \leftarrow S^{-1}\sum_{s}\ind\big(V^{(s)}_{ij}\ge V^*_{ij}\big)$ for every pair \tcp*{estimate of Eq.~\eqref{eq:pproj}}

\tcp{4. Multiple-testing correction}
\Return links and pairs selected by Benjamini--Hochberg at $\alpha_{\mathrm{sig}}$ (fixed threshold $0.05$ when BH selects none) \tcp*{Eq.~\eqref{eq:bh}}
\end{algorithm}

\clearpage

\section*{Data availability}
All datasets are publicly available from the repositories and original sources listed in Table~S1 of the Appendix. 

\section*{Code availability}
The WBNM package implementing fitting, sampling, model selection and validation is available at \github.

\section*{Acknowledgements}

\section*{Author contributions}
\textbf{Daniele Cirulli}: Software, Formal analysis, Investigation, Data Curation, Validation, Visualization, Writing - Original Draft;\\ 
\textbf{Lorenzo Buffa}: Software, Formal analysis, Investigation, Data Curation, Validation, Visualization, Writing - Original Draft;\\ 
\textbf{Giulio Cimini}: Conceptualization, Supervision, Writing - Review \& Editing;\\
\textbf{Dario Mazzilli}: Conceptualization, Data Curation, Supervision, Writing - Review \& Editing; \\
\textbf{Aurelio Patelli}: Conceptualization, Methodology, Data Curation, Supervision, Writing - Review \& Editing;\\
\textbf{Fabio Saracco}: Conceptualization, Methodology, Data Curation, Supervision, Writing - Review \& Editing, Project administration.

\section*{Competing interests}
The authors declare no competing interests.

\bibliographystyle{unsrtnat}

\bibliography{references}

\clearpage
\appendix

\let\oldsection\section\renewcommand{\section}{\FloatBarrier\oldsection}

\setcounter{section}{0}
\setcounter{figure}{0}
\setcounter{table}{0}
\setcounter{equation}{0}
\setcounter{algocf}{0}

\numberwithin{equation}{section}
\numberwithin{algocf}{section}
\numberwithin{figure}{section}
\numberwithin{table}{section}

\begin{center}{\LARGE\bfseries Appendix}\end{center}
\tableofcontents

\clearpage
\section{Model derivations and likelihoods}
\label{sec:S2}

\subsection{Dyadic distributions}
All models are maximum-entropy ensembles whose constraints are sums of dyadic quantities, so that the probability of a biadjacency matrix factorises over the dyads $(i,\alpha)$. With $\lambda_{i\alpha}=\Theta_i+H_\alpha$ the multipliers of the strengths, the weighted models have
\begin{align}
q^{\text{\biwcm}}_{i\alpha}(w)&=\lambda_{i\alpha}e^{-\lambda_{i\alpha}w},\qquad w\ge0,\\
q^{\text{hurdle}}_{i\alpha}(w)&=
\left\{
\begin{array}{ll}
1-p_{i\alpha} & w=0\\
p_{i\alpha}\,\lambda_{i\alpha}e^{-\lambda_{i\alpha}w} & w>0
\end{array}
\right.,
\end{align}
where the connection probability of the hurdle (topology-aware) models is
\begin{equation}
p^{\text{\bipecm}}_{i\alpha}=\frac{e^{-\delta}}{\lambda_{i\alpha}+e^{-\delta}},\qquad
p^{\text{\biecm}}_{i\alpha}=\frac{e^{-(\theta_i+\eta_\alpha)}}{\lambda_{i\alpha}+e^{-(\theta_i+\eta_\alpha)}},\qquad
p^{\text{\abiecm}}_{i\alpha}=\frac{e^{-(\theta_i+\eta_\alpha)}}{1+e^{-(\theta_i+\eta_\alpha)}}.
\end{equation}
Here $\delta$ is the multiplier of the total number of links $L$, and $\theta_i,\eta_\alpha$ those of the degrees $k_i,h_\alpha$. The \bicm is the purely binary model with $p^{\text{\bicm}}_{i\alpha}=p^{\text{\abiecm}}_{i\alpha}$. The qBiECM is discussed in Section~\ref{sec:qDECM}.

\subsection{Density log-likelihoods}
Summing $\ln q_{i\alpha}(w^*_{i\alpha})$ over all dyads at the maximum-likelihood solution gives, for the \bicm,
\begin{equation}
\ln\mathcal{L}_{\text{\bicm}}=-\sum_i\theta_ik_i^*-\sum_\alpha\eta_\alpha h_\alpha^*-\sum_{i,\alpha}\ln\left(1+e^{-(\theta_i+\eta_\alpha)}\right),
\label{eq:ll_bicm}
\end{equation}
which is already the log-probability of the observed binary matrix. For the weighted models the same sum is a log-density, which we denote by $\ell$:
\begin{align}
\ell_{\text{\biwcm}}&=-\sum_i\Theta_is_i^*-\sum_\alpha H_\alpha\sigma_\alpha^*+\sum_{i,\alpha}\ln\lambda_{i\alpha},
\label{eq:ll_biwcm}\\
\ell_{\text{\bipecm}}&=-\delta L^*-\sum_i\Theta_is_i^*-\sum_\alpha H_\alpha\sigma_\alpha^*-\sum_{i,\alpha}\ln\left(1+\frac{e^{-\delta}}{\lambda_{i\alpha}}\right),
\label{eq:ll_bipecm}\\
\ell_{\text{\biecm}}&=-\sum_i\theta_ik_i^*-\sum_\alpha\eta_\alpha h_\alpha^*-\sum_i\Theta_is_i^*-\sum_\alpha H_\alpha\sigma_\alpha^*-\sum_{i,\alpha}\ln\left(1+\frac{e^{-(\theta_i+\eta_\alpha)}}{\lambda_{i\alpha}}\right).
\label{eq:ll_biecm}
\end{align}
The \bicm depends only on the degree multipliers, the \biwcm only on the strength multipliers, the \bipecm combines the strength term with the single global multiplier $\delta$, and the \biecm combines strength and degree terms in full.

\subsection{Likelihood of the recorded data}
The densities above are dimensional quantities and cannot be compared with probabilities. We therefore compare models through the probability of the recorded values, obtained by integrating each continuous component over the observation bin $[w^*,w^*+\Delta w)$:
\begin{equation}
\Prob\{w^*\le W<w^*+\Delta w\}=e^{-\lambda w^*}\left(1-e^{-\lambda\Delta w}\right)=q(w^*)\,\frac{1-e^{-\lambda\Delta w}}{\lambda}.
\label{eq:binprob}
\end{equation}
The approximation $q(w^*)\Delta w$ of Ref.~\cite{cover2006elements} is recovered for $\lambda\Delta w\ll1$, but it is not a probability in general: for an empirical zero of the \biwcm it equals $\lambda\Delta w$, which exceeds one whenever the expected weight $1/\lambda$ is smaller than $\Delta w$, a frequent case at $\Delta w=1$ in sparse networks. With Eq.~\eqref{eqSI:binprob}, the log-likelihoods used in all tables and figures are
\begin{align}
\ln\mathcal{L}_{\text{\biwcm}}&=\ell_{\text{\biwcm}}+\sum_{i,\alpha}\ln\frac{1-e^{-\lambda_{i\alpha}\Delta w}}{\lambda_{i\alpha}},
\label{eqSI:ll_biwcm_corrected}\\
\ln\mathcal{L}_{\text{\bipecm}}&=\ell_{\text{\bipecm}}+\sum_{(i,\alpha):\,b^*_{i\alpha}=1}\ln\frac{1-e^{-\lambda_{i\alpha}\Delta w}}{\lambda_{i\alpha}}+\sum_{(i,\alpha):\,b^*_{i\alpha}=0}\ln\frac{1-p_{i\alpha}e^{-\lambda_{i\alpha}\Delta w}}{1-p_{i\alpha}},
\label{eqSI:ll_bipecm_corrected}\\
\ln\mathcal{L}_{\text{\biecm}}&=\ell_{\text{\biecm}}+\sum_{(i,\alpha):\,b^*_{i\alpha}=1}\ln\frac{1-e^{-\lambda_{i\alpha}\Delta w}}{\lambda_{i\alpha}}+\sum_{(i,\alpha):\,b^*_{i\alpha}=0}\ln\frac{1-p_{i\alpha}e^{-\lambda_{i\alpha}\Delta w}}{1-p_{i\alpha}}.
\label{eqSI:ll_biecm_corrected}
\end{align}
For the \biwcm the result is a geometric distribution over the bins, and an empirical zero has probability $1-e^{-\lambda_{i\alpha}\Delta w}$. For the hurdle models a recorded zero is either a structural zero or a link with weight below $\Delta w$, so that it has probability $(1-p_{i\alpha})+p_{i\alpha}(1-e^{-\lambda_{i\alpha}\Delta w})=1-p_{i\alpha}e^{-\lambda_{i\alpha}\Delta w}$, while a recorded link of weight $w^*$ has probability $p_{i\alpha}e^{-\lambda_{i\alpha}w^*}(1-e^{-\lambda_{i\alpha}\Delta w})$. Since $\ell$ already contains $\ln(1-p_{i\alpha})$ for each empty dyad, the last sum replaces it with $\ln(1-p_{i\alpha}e^{-\lambda_{i\alpha}\Delta w})$. With these corrections the probabilities of all recorded values sum to one in every model, and a recorded zero is treated in the same way, as the bin $[0,\Delta w)$, across the hierarchy.

The multipliers are the maximum-likelihood estimates of the continuous model, and the likelihood of the recorded data is evaluated at that solution. When $\lambda_{i\alpha}\Delta w\ll1$ the two likelihoods differ by a constant and share the same maximum; otherwise the values reported are lower bounds of the maximised likelihoods of the recorded data, for every model alike. The smallest BIC gap between the best and the second-best model is $88$ (Table~\ref{tab:model_selection}), so this does not affect model selection.

\subsection{Information criteria}
With $n=N_1N_2$ dyads and $k$ free parameters, $\mathrm{BIC}=k\ln n-2\ln\mathcal{L}$ and $\mathrm{AIC}=2k-2\ln\mathcal{L}$, with $k=N_1+N_2$ for the \biwcm (and the \bicm), $N_1+N_2+1$ for the \bipecm and $2(N_1+N_2)$ for the \biecm and the \abiecm. Gauge freedoms of the multipliers (one for each pair of multipliers entering only through a sum) are not subtracted; they shift $k$ by one or two units, which is negligible compared with the BIC gaps in Table~\ref{tab:model_selection}.

\section{Dataset sources and preprocessing}
\label{sec:S1}

We analyse 46 weighted bipartite matrices from seven systems (Table~\ref{tab:systems}): international trade, Reddit, a lexical corpus, Twitter, two families of technological matrices and ecological interaction networks. Repeated temporal snapshots are grouped under their parent system in every table. Basic statistics of each matrix are reported in Table~\ref{tab:input_network_statistics}. Rows and columns with zero strength are removed before any analysis.

\paragraph{Trade.} Country--product export matrices from UN Comtrade for the years 2012--2020 (biennial observations) \citep{JRC136943}. The data cover 220 countries and 5204 products classified at the 6-digit level of the Harmonized System 2012 (HS12). Each weight represents the export value of a country for a given product, expressed in constant 2015 US dollars.

\paragraph{Reddit.} Yearly matrices for 2013--2017, built from the Pushshift dumps~\cite{baumgartner2020pushshift,cirulli2025polarizationechochambersreddits}. Rows are subreddits, which are the nodes of the validated projections, and columns are news domains: a subreddit and a domain are linked if the domain was shared in at least one post of that subreddit. The weight of a link is the number of times the domain was shared in the subreddit.

\paragraph{Lexical corpus.}
The lexical corpus was downloaded from Ref.~\cite{Carungu2026}: it includes 599 sustainability or non-financial reports published from 2015 to 2023 by the 10 UK companies with the highest Company Market Cap listed on the London Stock Exchange. The preprocessing and the extraction of a weighted bipartite network between tokens and documents, in which weights encode the frequency of the various tokens in the different documents, follows a recipe similar to the one in Ref.~\cite{Dainelli2023}.
All texts under analysis were processed with the Natural Language Processing tools made available by the nltk Python library~\cite{Loper2002}. As a preprocessing step, individual tokens were extracted by stripping punctuation and purely numeric entries, while alphanumeric tokens (i.e.\ tokens containing both numbers and words) were retained; stop words were removed as well. Stop-word filtering relied on the English list provided in the nltk.corpus submodule~\cite{Loper2002}, which covers pronouns, demonstrative adjectives, prepositions, auxiliary verbs, and comparable function words.
The resulting tokens were then reduced to their root form using the Snowball stemmer~\cite{Loper2002}. Stemming strips prefixes and suffixes from a word, collapsing its inflected or derived forms onto a common stem, so that occurrences of the same underlying concept can be counted consistently.
Finally, the number of occurrences of each stem in each document was counted, providing a weighted biadjacency matrix in which rows are documents and columns are stems. Five yearly snapshots (2015, 2018, 2020, 2021, 2023) are analysed.

\paragraph{Twitter/X.} 
 We used tweets concerning the three most prominent political events debated on Italian Twitter/X in 2022, the same events analysed in Ref.~\cite{Guarino2026}.
Since, at the time of the download of the data, the platform name was still \emph{Twitter}, this will be the name of the dataset in the following.
For each event, a set of pertinent keywords was defined and used to retrieve tweets via Twitter's Academic API.
The three datasets comprise: 
\begin{itemize}
    \item the debate around the election of the President of the Republic (ended with the re-election of Sergio Mattarella on January 29, 2022).
    \item the debate around the political crisis that led to the fall of the Draghi government on July 21, 2022.
    \item the debate around the Italian general elections held on September 25, 2022, which led to Giorgia Meloni becoming Prime Minister.
\end{itemize}    
We emphasize that the data were collected \emph{prior to} the introduction of user verification as part of X's Premium subscription (\url{https://en.wikipedia.org/wiki/Twitter_verification}). 

For each dataset, we extracted from the collected tweets, following Refs.~\cite{Becatti2019d, Guarino2026}, a bipartite weighted network,
whose two layers are the verified users and the standard (i.e., non-verified) users. The weight of the link between a verified user $v$ and a standard user $s$ is a non-negative integer: the number of times $s$ reposted messages originally authored by $v$. Such an analysis can effectively extract users' political leanings; see Ref.~\cite{Becatti2019d,Guarino2026,Caldarelli2026}.

\paragraph{Technology.} 
\begin{itemize}
\item Country--technology patent matrices (CPC) for five biennial snapshots from 2010 to 2018. The data cover 47 major patent-producing countries and approximately 640 technology classes defined by the 4-digit Cooperative Patent Classification (CPC), restricted to sections A--H (excluding Y). Not all classes are active in every year. Weights represent fractional patent counts.
\item Country-level scientific citation matrices (CIT) for five biennial snapshots from 2010 to 2018, taken from \citep{PATELLI2023101357}. The data cover approximately 200 countries, with minor variations across years due to low publication activity in smaller countries. Countries are identified through authors' affiliations, and weights represent citations received by scientific papers, fractionally adjusted for the number of authors. The scientific classification accounts for 312 distinct scientific domains. The dataset is based on the Microsoft Academic Graph (MAG), whose legacy is now maintained within the OpenAlex project.

\end{itemize}

\paragraph{Ecological networks.} Mutualistic and host--parasite interaction matrices from IWDB and ICON~\cite{IWDB,ICON}. Weights are counts of visits or interactions (integers, $\Delta w=1$) or, for the host--parasite matrices of the Canadian lakes, parasite intensity and prevalence recorded with one decimal ($\Delta w=0.1$). Some matrices of the source collection coincide: Beehler~1 and Beehler~2 are identical, Snow~1 and Seed disp.\ Snow are the same matrix up to a permutation of rows and columns, and the intensity and prevalence matrices of McGregor lake are identical. For the other lakes, intensity and prevalence share the same support but have different weights. We keep a single copy of each duplicated matrix (Beehler, Seed disp.\ Snow and the McGregor intensity matrix), so that the corpus contains 46 distinct matrices.

\paragraph{Observation resolution.} All likelihoods are evaluated on the recorded values (Section~\ref{sec:S2}). For each matrix, the resolution $\Delta w$ is the largest power of ten such that every recorded weight is an integer multiple of it. Count data have $\Delta w=1$; ecological intensities and prevalences have $\Delta w=0.1$; trade values and fractional technology counts show no finite decimal resolution and are assigned $\Delta w=10^{-12}$. 

\input{SI_tabs/table_systems_v2}
\input{SI_tabs/table_network_inputs_v2}


\section{Structural drivers: which network properties favour the global-density model?}
\label{sec:str_dr}

The \bipecm and the \biecm differ only in the topological constraint: one global density against the full degree sequences. To identify the network properties that make the local constraints necessary, we compare the two models on the real networks and on an Erd\H{o}s--R\'enyi (ER) family in which the degrees carry no information beyond the density (Algorithm~\ref{alg:er_fixed_weight}, Section~\ref{sec:synth}).

\begin{figure}[t]
\centering
\includegraphics[width=\textwidth]{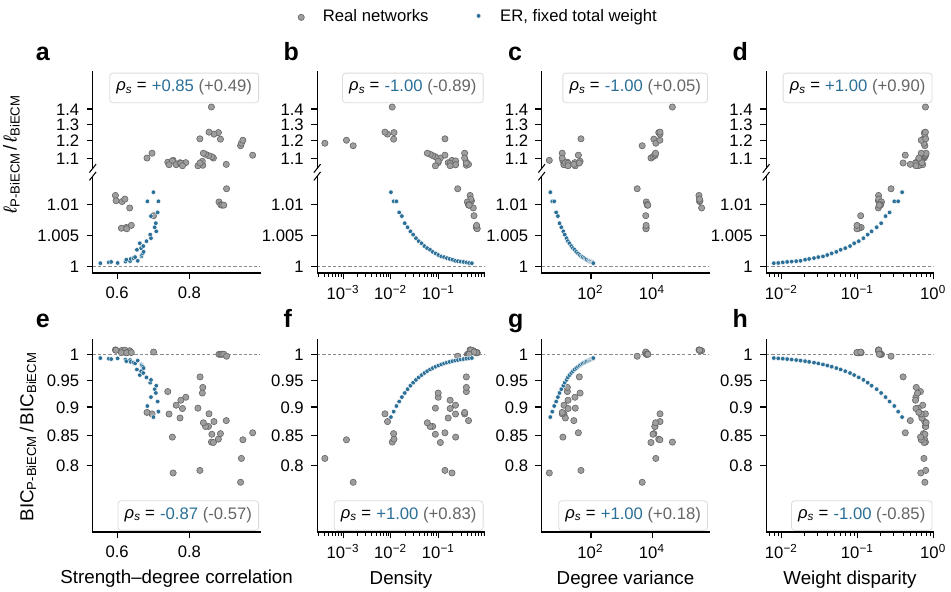}
\caption{\textbf{Global versus local topological constraints in synthetic and real networks.} \textbf{a--d},~Ratio between the log-likelihoods of the \bipecm and of the \biecm and \textbf{e--h},~ratio between their BIC values, against the strength--degree Spearman correlation (\textbf{a},\textbf{e}), the density (\textbf{b},\textbf{f}), the degree variance (\textbf{c},\textbf{g}) and the weight disparity (\textbf{d},\textbf{h}). Blue: ER networks at fixed total weight (Algorithm~\ref{alg:er_fixed_weight}; 30 densities, 3 realisations each). Grey: the empirical networks of Table~\ref{tab:input_network_statistics}. The boxes report the Spearman correlation for the ER family and, in parentheses, for the real networks. The vertical axis of \textbf{a--d} is broken to show both groups.}
\label{fig:er_ratios}
\end{figure}

Figure~\ref{fig:er_ratios} shows the ratios $\ell_{\text{\bipecm}}/\ell_{\text{\biecm}}$ of the log-likelihoods (both negative, so a ratio larger than one means that the \bipecm fits worse) and $\mathrm{BIC}_{\text{\bipecm}}/\mathrm{BIC}_{\text{\biecm}}$ (smaller than one when the \bipecm is selected). 

In the ER family, the likelihood ratio is always above one, since the \biecm has more parameters, but it approaches one as the density grows and the weight disparity decreases ($\rho_s=-1.00$ and $+1.00$), and the BIC ratio stays below one for every realisation: BIC recovers the generating global-density structure across the whole range of density.

Among the real networks, the same monotonic trends hold for density ($\rho_s=-0.89$ for the likelihood ratio and $+0.83$ for the BIC ratio) and weight disparity ($+0.90$ and $-0.85$), while the strength--degree correlation is a weaker predictor ($+0.49$ and $-0.57$) and the degree variance is not informative ($+0.05$ and $+0.18$).
Nonetheless, the ER family reproduces the sign of most of the trends observed in real networks (panels a, b, d, e, f, h), but not their magnitude: ER ensembles span likelihood and BIC ratios within about $10^{-2}$, whereas empirical networks reach up to 0.4, and occupy a region of the observables’ space that the ER family does not cover. The only observables showing a sizeable monotonic association with model preference in the empirical data are the density and the weight disparity (Spearman $|\rho_s| \approx 0.8–0.9$). However, the scatter around these trends is large (e.g. panel h, where at high disparity the BIC ratio spans roughly 0.77–1.0), so none of them yields a reliable \emph{a priori} criterion. BIC-based selection therefore remains necessary.

\section{Reconstruction and validation definitions}
\label{sec:reconstruction}

\subsection{Mesoscale observables}
With $b_{i\alpha}=\ind(w_{i\alpha}>0)$, degrees $k_i=\sum_\alpha b_{i\alpha}$, $h_\alpha=\sum_i b_{i\alpha}$ and strengths $s_i=\sum_\alpha w_{i\alpha}$, $\sigma_\alpha=\sum_i w_{i\alpha}$, the average nearest-neighbour degree and strength of a node of layer~1 are
\begin{equation}
k^{nn}_i=\frac{1}{k_i}\sum_\alpha b_{i\alpha}h_\alpha,\qquad s^{nn}_i=\frac{1}{k_i}\sum_\alpha b_{i\alpha}\sigma_\alpha,
\end{equation}
set to zero when $k_i=0$, and analogously for layer~2. Note that $s^{nn}_i$ is the average strength of the neighbours of $i$, not the weight-averaged nearest-neighbour degree.

\emph{Reconstruction errors (Table~\ref{tab:reconstruction_errors}).} For each network and model we sample $S=300$ networks from the fitted ensemble and compute $x^{(s)}_i$, $x\in\{k^{nn},s^{nn}\}$, for every node $i$ of a layer and every sample $s$. The ensemble mean and standard deviation of node $i$ are
\begin{equation}
\langle x_i\rangle=\frac1S\sum_{s=1}^S x^{(s)}_i,\qquad \sigma_{x_i}=\Big[\frac1S\sum_{s=1}^S\big(x^{(s)}_i-\langle x_i\rangle\big)^2\Big]^{1/2},
\end{equation}
and the deviation of the observed value is measured by the node z-score $z_i=(x^*_i-\langle x_i\rangle)/\sigma_{x_i}$, with $\sigma_{x_i}$ floored at $10^{-12}$. The table reports the root mean square of the z-scores over the $N$ nodes of the layer,
\begin{equation}
\mathrm{RMSE}_z=\Big[\frac{1}{N}\sum_{i=1}^N z_i^2\Big]^{1/2}.
\end{equation}
The root mean square is used instead of the average of $z_i$ because deviations of opposite sign would otherwise cancel. If the model is correct, each $z_i$ has zero mean and unit variance and $\mathrm{RMSE}_z\simeq1$; values much larger than one indicate deviations that the fluctuations of the model cannot account for. The unscaled error $\mathrm{RMSE}=[N^{-1}\sum_i(x^*_i-\langle x_i\rangle)^2]^{1/2}$ is computed as well and released with the parameter files; it is not comparable across networks because it carries the units of $x$. For the \bipecm and the \biecm on the Twitter matrices the $300$ samples are the first $300$ samples of the projection Monte Carlo (Section~\ref{sec:reconstruction}, validation of projections). Under the \biwcm every dyad is positive with probability one, so $k^{nn}_i$ is the same in every sample and its $\mathrm{RMSE}_z$ is not defined; $s^{nn}_i=T/N_2$ depends only on the total weight $T$ of the sample, and on Twitter $T$ is drawn from a normal distribution with the exact mean and variance of a sum of $N_1N_2$ independent exponentials instead of sampling the dense matrices.

\subsection{Macroscale observables}
\emph{Nestedness.} The binary NODF~\cite{AlmeidaNeto2008} is computed on each layer after ordering the nodes by decreasing degree: for each ordered pair $i<j$ the paired overlap is $N_{ij}=|\Gamma_i\cap\Gamma_j|/k_j$, the fraction of the neighbours of $j$ that are also neighbours of $i$. $N_{ij}$ is averaged over the pairs of each layer and $\mathrm{NODF}=50\,(\overline N_{\mathrm{rows}}+\overline N_{\mathrm{cols}})$, in percent. Unlike Ref.~\cite{AlmeidaNeto2008}, pairs with equal degree are not set to zero and the two layers have equal weight; the same definition is applied to the data and to every sample, so the comparison is unaffected. Its weighted counterpart is WNODA~\cite{Pinheiro2019}: nodes are ordered by decreasing strength, a pair with $s_i>s_j$ contributes the fraction of the positive entries of $j$ that are strictly smaller than the corresponding entries of $i$ (pairs with $s_i=s_j$ contribute zero), and the contributions are averaged over the pairs of both layers, in percent.

\emph{Clustering.} The binary bipartite clustering is the Robins--Alexander coefficient~\cite{RobinsAlexander2004} as implemented in NetworkX~\cite{Hagberg2008}, $C=4C_4/L_3$, with $C_4$ the number of 4-cycles and $L_3$ the number of paths of length three. The weighted clustering $C_W$ is the Opsahl coefficient~\cite{Opsahl2013}: the fraction of the total value of 4-paths $i$--$\alpha$--$j$--$\beta$--$l$ that are closed (i.e.\ $i$ and $l$ share a neighbour other than $\alpha,\beta$), with the value of a path given by the geometric mean of its four weights.

\emph{Representative networks (main-text reconstruction figure).} For Trade 2020 and Lake of the Woods we sample $S=2000$ networks from each model. At the mesoscale, besides $\mathrm{RMSE}_z$, we report the median relative error $\mathrm{med}_i|x^*_i-\langle x_i\rangle|/x^*_i$ and the fraction of nodes whose observed value lies within the central $95\%$ interval (2.5th to 97.5th percentile) of their sampled values; a correct model gives a fraction close to $0.95$. For a macroscale observable $y\in\{\mathrm{NODF},C,\mathrm{WNODA},C_W\}$, computed once per sample, we report the z-score $z=(y^*-\langle y\rangle)/\sigma_y$ and the percentile of $y^*$ in the sampled distribution, which is $0$ or $100$ when the observed value lies outside all samples. Results are in Table~\ref{tab:reconstruction_representative}.

\input{SI_tabs/table_reconstruction_representative_v2}

\subsection{Prediction of the empirical zeros}
\begin{figure}[t]
\centering
\includegraphics[width=0.9\textwidth]{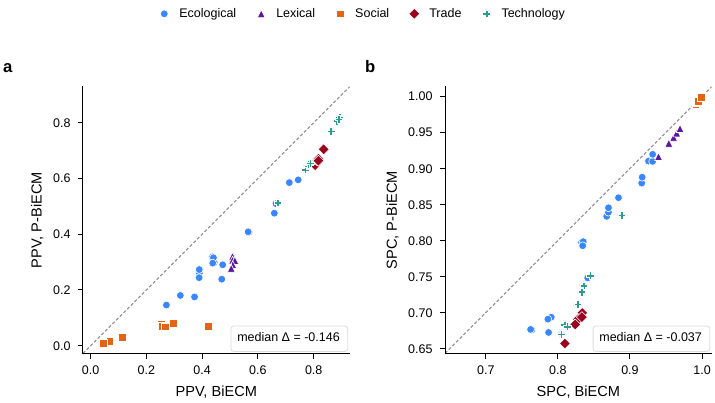}
\caption{\textbf{Prediction of the observed support by the topology-aware models.} \textbf{a},~Positive predictive value and \textbf{b},~specificity of the \bipecm against the \biecm for the empirical networks of Table~\ref{tab:input_network_statistics}. The dashed line is the identity; the boxes report the median difference \bipecm$-$\biecm.}
\label{fig:zero_prediction_si}
\end{figure}

For a topology-aware model with connection probabilities $p_{i\alpha}$, the expected confusion matrix with respect to the observed support $b^*_{i\alpha}$ is
\begin{equation}
\langle\mathrm{TP}\rangle=\sum_{b^*_{i\alpha}=1}p_{i\alpha},\quad
\langle\mathrm{FP}\rangle=\sum_{b^*_{i\alpha}=0}p_{i\alpha},\quad
\langle\mathrm{FN}\rangle=\sum_{b^*_{i\alpha}=1}(1-p_{i\alpha}),\quad
\langle\mathrm{TN}\rangle=\sum_{b^*_{i\alpha}=0}(1-p_{i\alpha}),
\end{equation}
from which $\mathrm{PPV}=\langle\mathrm{TP}\rangle/(\langle\mathrm{TP}\rangle+\langle\mathrm{FP}\rangle)$, the fraction of predicted links that are observed, and $\mathrm{SPC}=\langle\mathrm{TN}\rangle/(N_1N_2-L^*)$, the fraction of empirical zeros that are predicted as zeros. These are analytic expectations, not Monte Carlo estimates. The main text reports the PPV; Supplementary Fig.~\ref{fig:zero_prediction_si} reports both for the \bipecm and the \biecm.

\subsection{Validation of bipartite weights}
For an observed positive weight, the upper-tail probability under a hurdle model is $p^{\mathrm{link}}_{i\alpha}=p_{i\alpha}\exp(-\lambda_{i\alpha}w^*_{i\alpha})$, with $p_{i\alpha}=1$ for the \biwcm. 
\subsection{Validation of projections}
For two nodes $i,j$ of layer~1 the empirical co-volume is $V^*_{ij}=\sum_\alpha w^*_{i\alpha}w^*_{j\alpha}$. Its null distribution is estimated from $S$ networks sampled from the fitted ensemble, and the p-value
$p^{\mathrm{proj}}_{ij}=\Prob(V^{\mathrm{null}}_{ij}\ge V^*_{ij})$ is estimated as
\begin{equation}
\widehat p_{ij}=\frac{1}{S}\sum_{s=1}^{S}\ind\big(V^{(s)}_{ij}\ge V^*_{ij}\big).
\end{equation}
The real networks use $S=10^3$, except plants--ants ($S=10^6$) and the Twitter matrices, whose Benjamini--Hochberg thresholds are below $10^{-2}$ and are not resolved with $10^3$ samples: for the \bipecm and the \biecm we use $S=3\times10^4$ (Q22) and $S=10^5$ (election, crisis), accumulated over independent runs with distinct seeds. A threshold $t_{\mathrm{BH}}$ is resolved when $t_{\mathrm{BH}}S\ge10$ exceedances are expected; this holds for every network and model except the \biecm on election and crisis ($t_{\mathrm{BH}}S\simeq2$), whose validated sets may still shrink with more samples. The number of samples of every network is given in Table~\ref{tab:projection_characteristics}. Pairs are validated with the Benjamini--Hochberg procedure at $\alpha_{\mathrm{sig}}=0.05$; when it does not reject any hypothesis for at least one model, all models of that network are validated with the common threshold $\widehat p_{ij}<0.05$, so that the sets being compared are obtained with the same rule (column Thr.\ of Table~\ref{tab:projection_characteristics}). This happens for all ecological matrices, whose small size makes the FDR procedure too conservative.

For the Twitter matrices (crisis: $6667\times321824$) a single \biwcm sample is complete and too expensive to generate in the numbers needed to resolve their thresholds. Since the \biwcm weights are independent exponentials with means $\mu_{i\alpha}=1/\lambda_{i\alpha}$, the cumulants of $V_{ij}$ are exact,
\begin{equation}
\kappa_1=\sum_\alpha\mu_{i\alpha}\mu_{j\alpha},\quad \kappa_2=3\sum_\alpha\mu^2_{i\alpha}\mu^2_{j\alpha},\quad \kappa_3=26\sum_\alpha\mu^3_{i\alpha}\mu^3_{j\alpha},\quad \kappa_4=426\sum_\alpha\mu^4_{i\alpha}\mu^4_{j\alpha},
\end{equation}
and the \biwcm p-values are obtained from the Edgeworth expansion of the tail to fourth order. On the election matrix, where the Monte Carlo estimate is available, the two procedures give the same FDR threshold and an F1 overlap of 0.9998 between the validated sets.

\emph{Monte Carlo resolution.} With $S$ samples the smallest non-zero p-value is $1/S$, and the standard error of $\widehat p_{ij}$ is $[p^{\mathrm{proj}}_{ij}(1-p^{\mathrm{proj}}_{ij})/S]^{1/2}$ (0.007 at $p^{\mathrm{proj}}_{ij}=0.05$ for $S=10^3$), so the fixed threshold $0.05$ is well resolved with $S=10^3$. The Benjamini--Hochberg thresholds are much smaller on the largest matrices, and the number of samples is increased until they are resolved ($t_{\mathrm{BH}}S\ge10$), with the two exceptions given above.


\section{Complete empirical results}
\label{sec:S6}

Tables~\ref{tab:model_selection}, \ref{tab:reconstruction_errors} and \ref{tab:projection_characteristics} report, for every network of the corpus, the model comparison, the reconstruction errors and the comparison of the validated projections. Model names are the same in every table. BIC selects the \bipecm in 32 networks and the \biecm in 14 (all trade and Tech CIT matrices and the Tech CPC matrices up to 2016); AIC, which penalises parameters less, selects the \biecm in 40 networks. The \biwcm is never selected.

\input{SI_tabs/table_model_selection_v3}

\input{SI_tabs/table_reconstruction_errors_v2}

\input{SI_tabs/table_projection_characteristics_v3}



\section{Network properties and projection disagreement}\label{sec:io}
This analysis is exploratory. Its purpose is to see which properties of the input networks appear to accompany the differences between the validated projections of the \biwcm, the \bipecm and the \biecm, and to guide the controlled experiments of Section~\ref{sec:synth}; it is not meant to establish causal effects. For comparability, all pairs are validated here with the common threshold $\widehat p<0.05$ for every network and model, so the validated fractions and F1 overlaps of the non-ecological networks differ from those of Table~\ref{tab:projection_characteristics}, which use the Benjamini--Hochberg procedure.

For each of the 46 networks we consider nine outcomes: the Spearman correlation between the p-values of each pair of models, which does not depend on any threshold, the F1 overlap between the validated sets of each pair and the validated fraction of each model. The F1 overlap is not defined when neither model validates any pair (three ecological networks for the pairs including the \biwcm, five for the \bipecm--\biecm pair), and these networks are excluded from the corresponding tests. We use twelve descriptors: the five used for the synthetic families (density $d$ and mean degree $\langle k\rangle$, as in Table~\ref{tab:input_network_statistics}, the variance of the degrees of both layers pooled, and the mean and variance of the non-zero weights) and seven further ones (the layer sizes $N_1$ and $N_2$, the mean binary co-occurrence $\bar V=\sum_{i\neq j}\sum_\alpha b_{i\alpha}b_{j\alpha}/[N_1(N_1-1)]$, the coefficients of variation of the degrees and of the weights, the strength--degree Spearman correlation and the weight disparity). Mean weight and weight variance depend on the units of each system and are compared across systems only indicatively.

We use only non-parametric tests: Spearman correlations across networks with Benjamini--Hochberg correction over the 108 descriptor--outcome pairs (adjusted p-values $p_{\mathrm{BH}}$), and, for the five descriptors of the synthetic families, Kruskal--Wallis tests across terciles with effect size $\eta^2_H=(H-k+1)/(n-k)$. Since snapshots of the same system are not independent, we also report a within-system permutation test, in which the outcome is permuted among networks of the same system ($10^4$ permutations; adjusted p-values $p^{\mathrm{within}}_{\mathrm{BH}}$), and a 95\% bootstrap confidence interval (CI) over the seven systems ($5\times10^3$ resamples).

\begin{figure}[t]
\centering
\includegraphics[width=\textwidth]{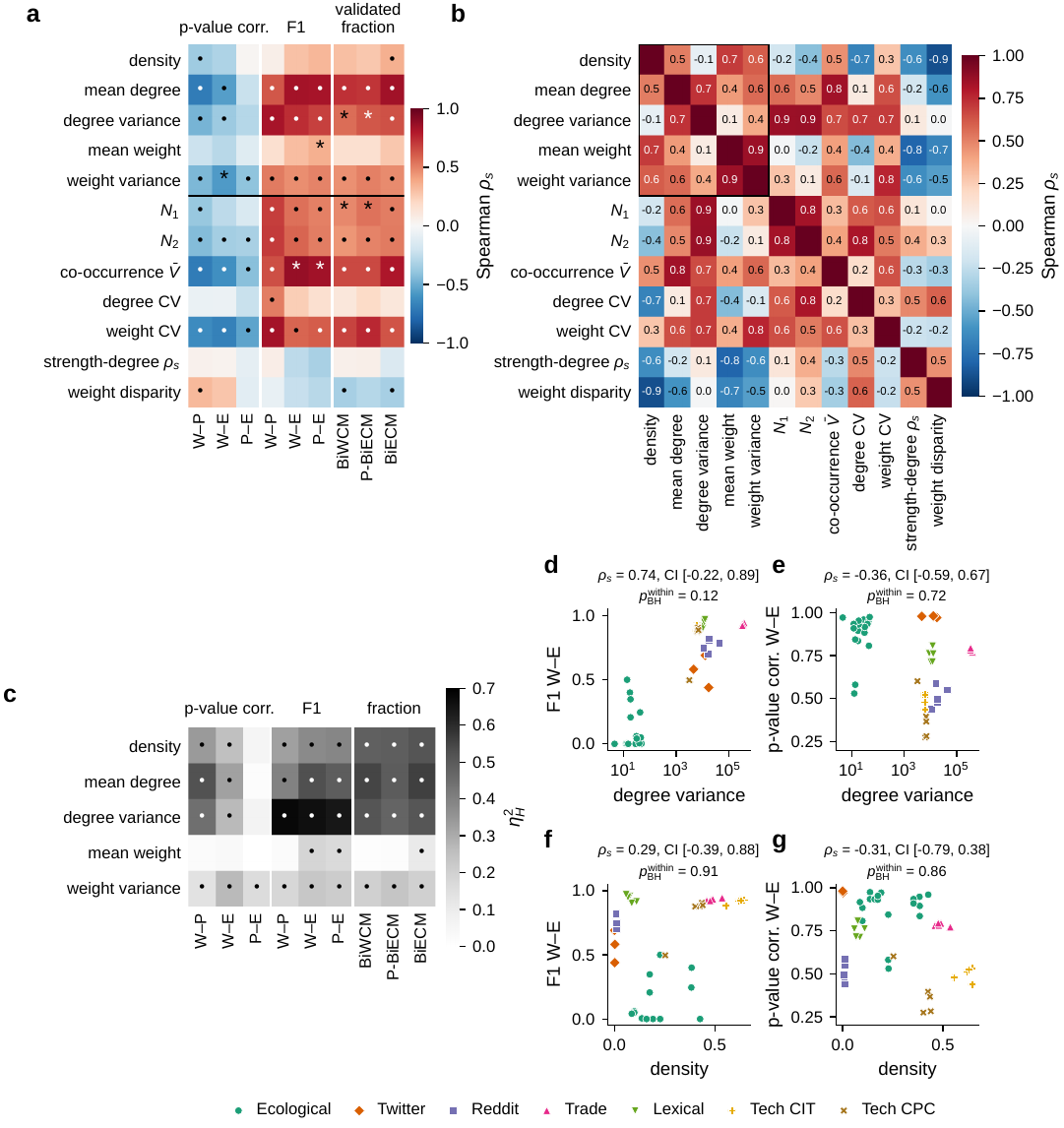}
\caption{\textbf{Input-network properties and projection disagreement in the 46 empirical networks (exploratory).} All pairs are validated with $\widehat p<0.05$. \textbf{a},~Spearman correlation between twelve input descriptors (rows; above the line, the five descriptors of the synthetic families) and nine projection outcomes: the Spearman correlation between the p-values and the F1 overlap of the validated sets for each pair of models (W: \biwcm, P: \bipecm, E: \biecm), and the validated fraction of each model. Dots: $p_{\mathrm{BH}}<0.05$ over the 108 tests; asterisks: also $p^{\mathrm{within}}_{\mathrm{BH}}<0.05$ in the within-system permutation test. \textbf{b},~Spearman correlation between the descriptors; the five descriptors of the synthetic families are framed. \textbf{c},~Kruskal--Wallis effect size $\eta^2_H$ across the terciles of these five descriptors; dots: $p_{\mathrm{BH}}<0.05$. \textbf{d}--\textbf{g},~F1 overlap (\textbf{d},\textbf{f}) and p-value correlation (\textbf{e},\textbf{g}) between the \biwcm and the \biecm against the degree variance (\textbf{d},\textbf{e}) and the density (\textbf{f},\textbf{g}), coloured by system. Above each panel: Spearman correlation $\rho_s$, 95\% CI and $p^{\mathrm{within}}_{\mathrm{BH}}$ of the within-system permutation test. }
\label{fig:si_input_output}
\end{figure}

Many descriptors are correlated with the outcomes across the corpus (66 of 108 pairs with $p_{\mathrm{BH}}<0.05$,, Fig.~\ref{fig:si_input_output}a), but only a few of these associations persist within systems (eight), so most of them reflect differences between systems. We therefore read the results only as indications.

\vspace{-0.15cm}

\begin{itemize}
\item \emph{Degree variance.} Among the five descriptors of the synthetic families, it is the one that appears most associated with the outcomes (mean $\eta^2_H=0.48$, against 0.43 for the mean degree, 0.36 for density, 0.20 for the weight variance and 0.06 for the mean weight; Fig.~\ref{fig:si_input_output}c). Networks with larger degree variance tend to have higher F1 overlaps ($\rho_s$ between $0.68$ and $0.82$) and lower correlations between p-values ($\rho_s$ between $-0.47$ and $-0.28$; Fig.~\ref{fig:si_input_output}d,e). A possible reading is that, when degrees are heterogeneous, the models validate similar sets of pairs but rank them differently. In the real networks the degree variance also grows with size (Fig.~\ref{fig:si_input_output}b), so its role cannot be separated here; this is one motivation for the degree-heterogeneity family of Section~\ref{sec:synth}.
\item \emph{Density.} The associations are weak ($|\rho_s|\le0.35$) and not monotonic: the middle tercile, mostly ecological networks, has the lowest validated fractions (median 0.14 for the \biwcm against 0.83 and 0.95 in the other terciles; Fig.~\ref{fig:si_input_output}f,g). In the real networks density does not appear to be a good indicator of projection disagreement.
\item \emph{Co-occurrence.} The mean co-occurrence $\bar V$ shows the clearest association, the only one that persists both within systems and in the bootstrap over systems: it is correlated with the F1 overlap of the \biwcm--\biecm ($\rho_s=0.87$, 95\% CI $[0.54,0.95]$) and \bipecm--\biecm pairs ($0.84$, $[0.39,0.94]$). Networks with more co-occurrences per pair seem to provide more evidence for every model, so that more pairs are validated by all of them.
\item \emph{Weight descriptors.} Mean weight, weight variance, strength--degree correlation and weight disparity show no clear pattern.
\end{itemize}

\section{Synthetic network experiments}
\label{sec:synth}

We use three synthetic families, all with $N_1=N_2=500$. The ER family at fixed total weight (Algorithm~\ref{alg:er_fixed_weight}) tests model selection between global and local topological constraints (Section~\ref{sec:str_dr}). The transition family (Algorithm~\ref{alg:transition}) and the degree-heterogeneity family (Algorithm~\ref{alg:degree_heterogeneity}) test how the information encoded in topology and weights changes the validated projections. All networks of a family are fitted with the \bicm, the \biwcm, the \bipecm and the \biecm. The transition family moves the block structure from the links to the weights, testing the hierarchy of the null models (topology only, strengths only, or both~\cite{Mastrandrea2014}); the degree-heterogeneity family varies the degree variance, which appears associated with projection disagreement in the real networks (Section~\ref{sec:io}). Along the transition density, weights and degrees change together, while the degree-heterogeneity family keeps density and weights fixed; in both, only the direction of the effects is compared with the real networks.

\begin{algorithm}[H]
\small
\DontPrintSemicolon
\caption{Erd\H{o}s--R\'enyi family at fixed total weight.}
\label{alg:er_fixed_weight}
\KwIn{$N_1=N_2=500$; grid $p\in[10^{-2},0.5]$ (30 log-spaced values); total weight $W_{\mathrm{tot}}=10^{7}$; 3 independent realisations}
draw $u_{i\alpha}\sim\mathcal{U}(0,1)$ once per realisation\;
\For{each $p$}{
  $b_{i\alpha}\gets\ind[u_{i\alpha}<p]$, \quad $L\gets\sum_{i,\alpha}b_{i\alpha}$\;
  draw shares $g_\ell\sim\mathrm{Gamma}(1,1)$ for the $L$ links, \quad $(m_\ell)\sim\mathrm{Multinomial}\big(W_{\mathrm{tot}}-L;\ g_\ell/\textstyle\sum_\ell g_\ell\big)$\;
  $w_\ell\gets1+m_\ell$ on each link, so that $\sum_{i,\alpha}w_{i\alpha}=W_{\mathrm{tot}}$ with integer weights ($\Delta w=1$)\;
}
\end{algorithm}

\begin{algorithm}[H]
\small
\DontPrintSemicolon
\caption{Transition family: the block structure moves from the topology to the weights.}
\label{alg:transition}
\KwIn{$N_1=N_2=500$, two equal blocks; grid $\delta\in[10^{-3},1]$ (50 log-spaced values); one realisation per value}
\For{each $\delta$}{
  $p_{\mathrm{in}}\gets0.30+0.70\,\delta$, \quad $p_{\mathrm{out}}\gets0.01+0.99\,\delta$\;
  $w_{\mathrm{in}}\gets10^{2}+(10^{4}-10^{2})\,\delta$, \quad $w_{\mathrm{out}}\gets10^{2}-(10^{2}-1)\,\delta$\;
  \For{each pair $(i,\alpha)$}{
    $(p,w)\gets(p_{\mathrm{in}},w_{\mathrm{in}})$ if $i$ and $\alpha$ are in the same block, else $(p_{\mathrm{out}},w_{\mathrm{out}})$\;
    $b_{i\alpha}\sim\mathrm{Bernoulli}(p)$, \quad $w_{i\alpha}\gets b_{i\alpha}\,u_{i\alpha}$, \quad $u_{i\alpha}\sim\mathcal{U}(0.9\,w,\ 1.1\,w)$\;
  }
}
\end{algorithm}

\begin{algorithm}[H]
\small
\DontPrintSemicolon
\caption{Degree-heterogeneity family at fixed density and block contrast.}
\label{alg:degree_heterogeneity}
\KwIn{$N_1=N_2=500$, two equal blocks; target densities $p_{\mathrm{in}}=0.70$, $p_{\mathrm{out}}=0.20$; grid $\delta\in\{0,0.025,\dots,0.45\}$ (19 values); 3 independent realisations}
draw $z_i,z_\alpha\sim\mathcal{N}(0,1)$ once per realisation\;
\For{each $\delta$}{
  target degrees $\kappa_i\propto e^{\delta z_i}$, $\kappa_\alpha\propto e^{\delta z_\alpha}$, rescaled to the mean degree implied by $p_{\mathrm{in}},p_{\mathrm{out}}$\;
  $p_{i\alpha}\gets\dfrac{x_i\,y_\alpha\,c_{i\alpha}}{1+x_i\,y_\alpha\,c_{i\alpha}}$, \quad $c_{i\alpha}=1$ within blocks, $c_{i\alpha}=c_{\mathrm{out}}$ across blocks\;
  solve for $x_i,y_\alpha,c_{\mathrm{out}}$ such that $\sum_\alpha p_{i\alpha}=\kappa_i$, $\sum_i p_{i\alpha}=\kappa_\alpha$ and the expected density across blocks equals $p_{\mathrm{out}}$\;
  $b_{i\alpha}\sim\mathrm{Bernoulli}(p_{i\alpha})$, \quad $w_{i\alpha}\gets b_{i\alpha}\,u_{i\alpha}$, \quad $u_{i\alpha}\sim\mathcal{U}(0.5\cdot10^{5},\ 1.5\cdot10^{5})$\;
}
\end{algorithm}

\paragraph{ER family.} The topology is homogeneous at every density, so the degrees carry no information beyond $L$ and the global-density model is the correct topological description. Fixing the total weight while increasing the density spreads the same volume over more links, so density and weight disparity vary together. The family therefore checks whether BIC recovers the global-density model (it does in all 90 networks, Fig.~\ref{fig:er_ratios}) and how close to the decision boundary the likelihood ratio gets as the network densifies.

\paragraph{Transition family.} At small $\delta$ the two blocks are encoded in the topology ($p_{\mathrm{in}}=0.30$, $p_{\mathrm{out}}=0.01$) with nearly uniform weights; at $\delta=1$ the network is complete and the blocks are encoded only in the weights ($10^4$ against $1$). The family is therefore an interpolation between a block structure carried by the binary links and one carried only by the weights.

\paragraph{Degree-heterogeneity family.} Degree heterogeneity increases with $\delta$ while the expected density within and across blocks is kept fixed: the connection probabilities have the \bicm form with a block factor, so they remain below one for every $\delta$ and neither the density nor the block contrast changes with heterogeneity. Weights are nearly uniform, so the strengths follow the degrees. In the figures we use $\delta\le0.325$: beyond this value the F1 overlap between the \bicm and the \biwcm no longer changes beyond the variability between realisations. A previous version of this family, with multiplicative lognormal propensities $p_{i\alpha}=\min(1,p_{\mathrm{block}}\alpha_i\beta_\alpha)$, was discarded because the truncation at one made density and block contrast decrease together with $\delta$.

\paragraph{Validation and sample size.} For each network and model, the projection p-values are estimated from $S=10^3$ sampled networks, and pairs are validated with the fixed threshold $\widehat p<0.05$, the same for all models, so that the overlap between validated sets reflects the null models and not different thresholds. The threshold is fifty times the p-value resolution $1/S$, and the Monte Carlo standard error at $p^{\mathrm{proj}}_{ij}=0.05$ is $0.007$; with $S=10^3$ the validated sets are therefore stable, while the Benjamini--Hochberg thresholds can fall below $1/S$ and would not be resolved. Each figure reports, besides the F1 overlap, the Pearson correlation between the p-values of all pairs, which does not depend on any threshold.

\begin{figure}[t]
\centering
\includegraphics[width=\textwidth]{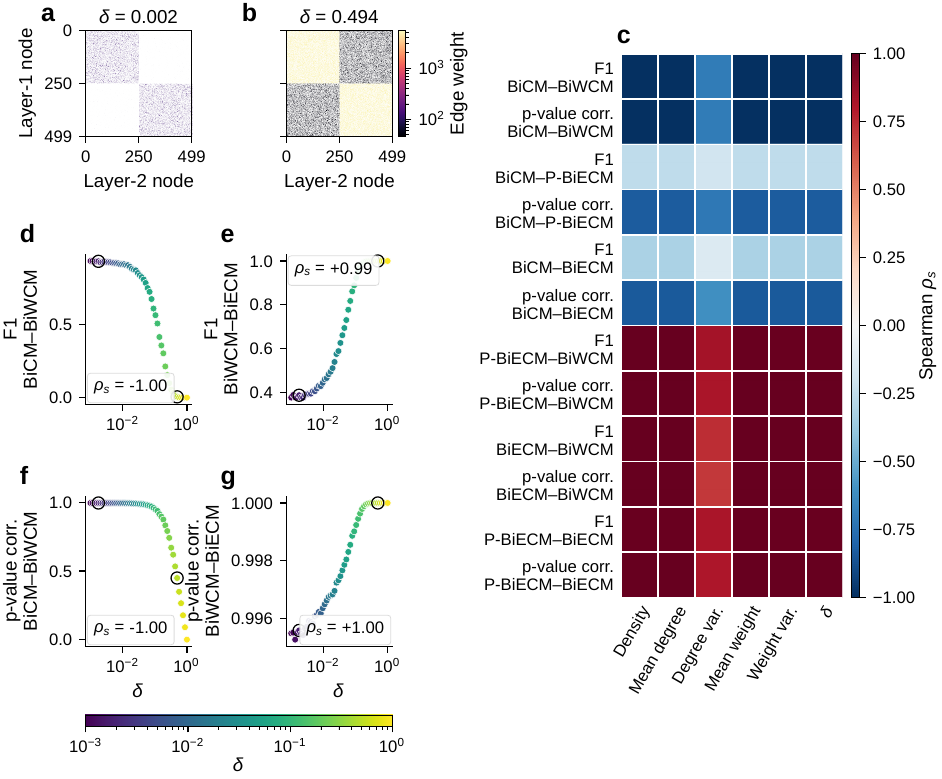}
\caption{\textbf{Null-model agreement across a topology-to-weight interpolation} (Algorithm~\ref{alg:transition}). \textbf{a},\textbf{b},~Weighted biadjacency matrices at low and high $\delta$ (white: no link). \textbf{c},~Spearman correlation between input-network properties and two agreement measures between the projections of each pair of models: the F1 overlap between the validated sets and the Pearson correlation between the p-values. \textbf{d},\textbf{e},~F1 overlap between the \bicm and the \biwcm and between the \biwcm and the \biecm against $\delta$. \textbf{f},\textbf{g},~Pearson correlation between the p-values of all pairs for the same two pairs of models. Colour encodes $\delta$, and the circled markers are the networks of \textbf{a} and \textbf{b}. Pairs are validated with $\widehat p<0.05$ for all models, with $10^3$ samples per network and model.}
\label{fig:transition}
\end{figure}

\paragraph{Results.} In the transition family (Fig.~\ref{fig:transition}) the agreement between the \bicm and the \biwcm falls from $\mathrm{F1}\simeq0.93$ to zero as $\delta$ grows ($\rho_s=-1.00$), while the agreement between the \biwcm and the \biecm rises from $0.38$ to one ($\rho_s=+0.99$): as the structure moves into the weights, the topology-only model loses the signal and the topology-aware weighted model converges to the strength-only one, because the network becomes complete. The same pattern holds for the p-value correlations and for all pairs involving the \bipecm (panel~c).

\begin{figure}[t]
\centering
\includegraphics[width=\textwidth]{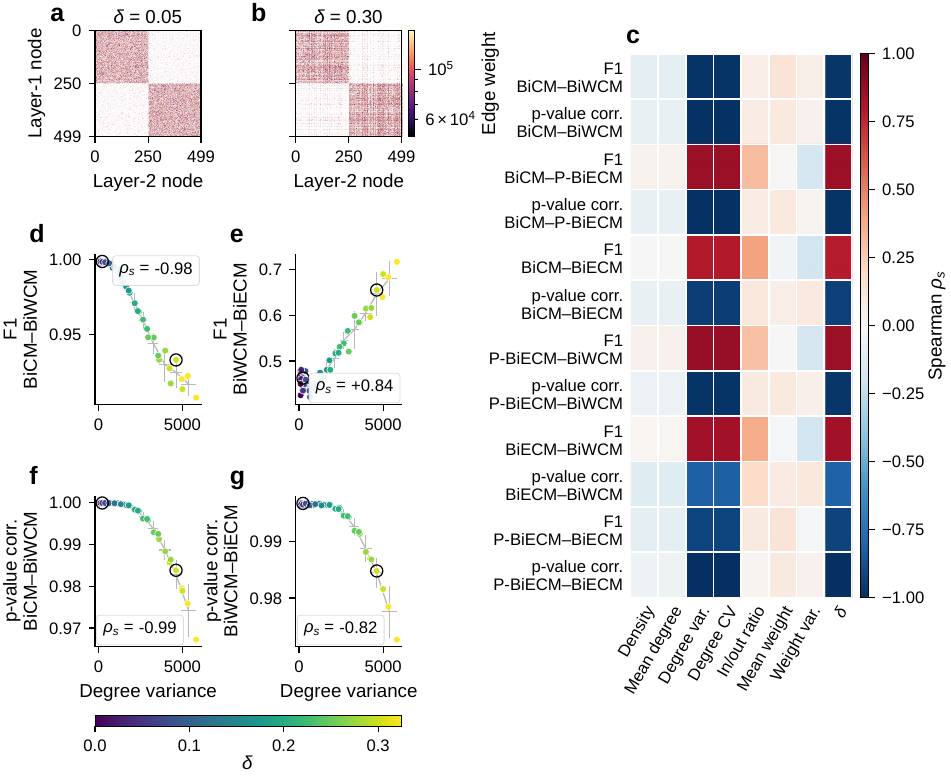}
\caption{\textbf{Null-model agreement under increasing degree heterogeneity at fixed density} (Algorithm~\ref{alg:degree_heterogeneity}). \textbf{a},\textbf{b},~Weighted biadjacency matrices at low and high $\delta$. \textbf{c},~Spearman correlation between input-network properties and the agreement measures between pairs of models, as in Fig.~\ref{fig:transition}. \textbf{d},\textbf{e},~F1 overlap between the \bicm and the \biwcm and between the \biwcm and the \biecm against the degree variance, for three realisations per $\delta$. \textbf{f},\textbf{g},~Pearson correlation between the p-values of all pairs for the same two pairs of models: unlike the F1 overlap in \textbf{e}, it decreases for both pairs. Grey crosses are means and standard deviations over realisations at fixed $\delta$, and the circled markers are the networks of \textbf{a} and \textbf{b}. Pairs are validated with $\widehat p<0.05$ for all models, with $10^3$ samples per network and model. }
\label{fig:degree_heterogeneity}
\end{figure}

In the degree-heterogeneity family (Fig.~\ref{fig:degree_heterogeneity}) the two agreement measures behave differently, unlike in the transition family where both increase together. The F1 overlap between the \bicm and the \biwcm decreases from $1.00$ to about $0.91$ as the degree variance grows ($\rho_s=-0.98$), and the Pearson correlation between their p-values decreases accordingly (from $1.000$ to $0.97$ at $\delta=0.325$). The F1 overlap between the \biwcm and the \biecm instead increases, from about $0.45$ to $0.68$ ($\rho_s=+0.84$), while the correlation between their p-values decreases, from $0.996$ to $0.978$ (to $0.917$ at $\delta=0.45$). The two observations are consistent: the pairs validated by the \biecm are almost all (95--100\%) also validated by the \biwcm, and as heterogeneity grows the \biwcm validates fewer pairs (from 50\% to 41\% of all pairs) and the \biecm more (from 15\% to 21\%). For nested sets the F1 overlap is $2b/(a+b)$, with $a$ and $b$ the two validated fractions, so it increases when the two sizes become closer, even if the p-values of the two models become less correlated. The increase of the F1 overlap therefore does not indicate a convergence of the two null models: the threshold-independent p-value correlation shows that agreement decreases with degree heterogeneity, for both pairs (panels~f and g) and for all pairs of models (panel~c). The same decrease of the p-value correlation is found in the sparse variant of the family ($p_{\mathrm{in}}=0.20$, $p_{\mathrm{out}}=0.05$, from $0.89$ to $0.49$) and in the previous version of the family. Density, mean degree and mean weight are uncorrelated with all the agreement measures (panel~c), confirming that the controlled variable is heterogeneity alone. 
In this dense family the Spearman correlation between the p-values stays close to $0.92$ for both pairs, so heterogeneity changes the values of the p-values more than their ranking; in the sparse variant the Spearman correlation decreases as well, from $0.96$ to $0.48$.

\section{Approximate enhanced inference: qBiECM}
\label{sec:qDECM}

Solving the parameters of the \biecm can be computationally demanding. Therefore, in our toolkit, we also provide the quasi-Bipartite Enhanced Configuration Model (\abiecm), a method inspired by the \biecm that introduces a simplification in the derivation and reduces the computation time of the parameters\footnote{The monopartite version of the \abiecm was introduced independently by Refs.~\cite{Gabrielli2019} and \cite{Parisi2020}, respectively called \emph{Separable-ECM} and \emph{Conditional Reconstruction Method} (CReM). We adopt a different name because neither is fully appropriate in our setting. The model is not truly \emph{separable}, since solving the topology independently of the weights is only justified under specific conditions (see Ref.~\cite{Gabrielli2019}). Likewise, we are not reconstructing a network from partial information, but describing a statistical model. The term ``quasi'' reflects the fact that the two-step maximisation of the Shannon entropy does not follow the standard procedure, while retaining part of its philosophy.}.

The maximisation of the entropy is split into two steps. First, a \bicm~\cite{Saracco2015} is fitted to the empirical degrees, which gives the connection probabilities $p^{\text{\abiecm}}_{i\alpha}=e^{-(\theta_i+\eta_\alpha)}/[1+e^{-(\theta_i+\eta_\alpha)}]$, independent of the strength multipliers. Second, the exponential weight kernel is fitted to the strengths conditionally on these probabilities, i.e.\ by imposing $\sum_\alpha p_{i\alpha}/\lambda_{i\alpha}=s^*_i$ and $\sum_i p_{i\alpha}/\lambda_{i\alpha}=\sigma^*_\alpha$. The resulting dyadic law has the same hurdle form as the \biecm. The approximation neglects the dependence of the topology on the weights and is not the joint maximum-entropy solution; for this reason the \abiecm is not a member of the hierarchy used for model selection. Its log-likelihood, evaluated with the same expressions as the \biecm, can exceed that of the \biecm, because the two models are not optimised over the same parametrised family.

The \abiecm converges on all networks with the coordinate solver, at a cost similar to or lower than that of the \biecm (Section~\ref{sec:numerical}).


\section{Numerical implementation and convergence}
\label{sec:numerical}

The complete workflow is implemented in an open-source Python package built on PyTorch, exposing a common interface for the null models (\bicm, \biwcm, \bipecm, \biecm and \abiecm) and complementing existing maximum-entropy solvers such as NEMtropy~\cite{Vallarano2021}. All solvers are vectorised over dyads, can be chunked row-wise to bound memory use on large matrices, and run on GPU when requested.

Parameter fitting solves the moment-matching equations $\langle\bm C\rangle=\bm C^*$. Three solvers are available.
\begin{itemize}
\item \emph{Fixed point.} The multipliers are updated multiplicatively with the relative constraint error, $\bm x^{(t+1)}=\bm x^{(t)}\odot(1+\gamma\bm v^{(t)})$, where $\bm v$ is a Polyak-style momentum average of $\langle\bm C\rangle/\bm C^*-1$ (step $\gamma=1$, momentum $\beta=0.9$, optionally annealed to $0.999$ during a warm-up window). The \biwcm converges without momentum.
\item \emph{Quasi-Newton.} A diagonal approximation of the Jacobian of the constraints is used as preconditioner, with a model-specific step size.
\item \emph{Coordinate.} Each multiplier is updated in turn by solving its own equation with the others fixed (bisection in log-space, 60 steps), which guarantees monotonic convergence when the fixed point is unstable. 
\end{itemize}
Convergence is assessed on the maximum relative error between empirical and expected constraints, $\max|\langle\bm C\rangle/\bm C^*-1|$: in all the analyses of this work a solution is accepted when this error is below $10^{-3}$.

\begin{figure}[t]
\centering
\includegraphics[width=\textwidth]{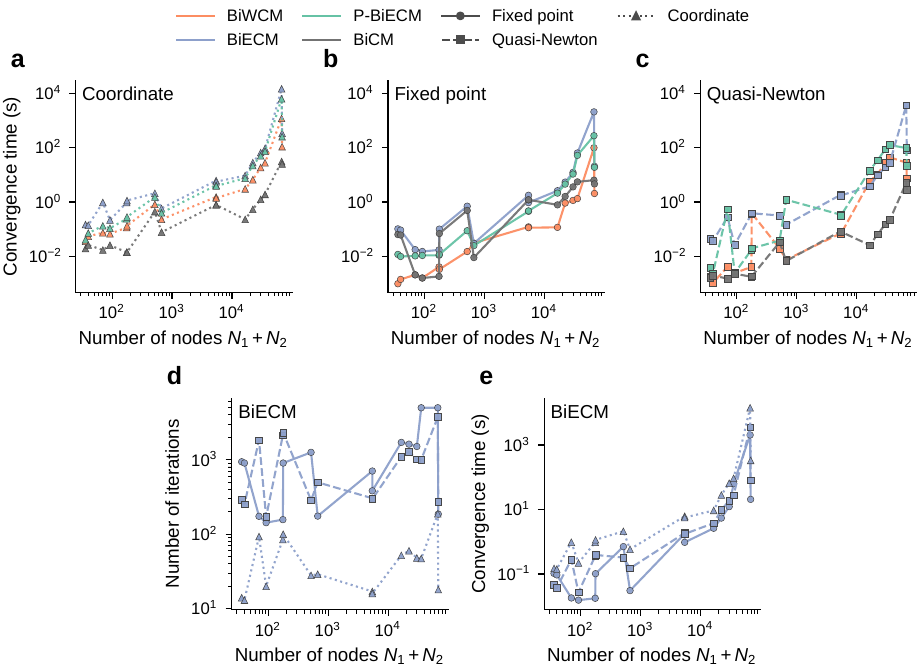}
\caption{\textbf{Computational performance of the fitting solvers.} \textbf{a--c},~Convergence time against the number of nodes $N_1+N_2$ for the \biwcm, the \biecm, the \bipecm and the \bicm, with the coordinate (\textbf{a}), fixed-point (\textbf{b}) and quasi-Newton (\textbf{c}) solvers. \textbf{d},\textbf{e},~Number of iterations (\textbf{d}) and convergence time (\textbf{e}) of the \biecm for the three solvers. In each panel the networks are selected among those for which all the plotted model--solver pairs are available, taking up to two networks per logarithmic interval of $N_1+N_2$ (10 intervals). Statistics over all networks are in Table~\ref{tab:benchmark}.}
\label{fig:benchmark}
\end{figure}

\paragraph{Benchmark.} Figure~\ref{fig:benchmark} shows convergence time and number of iterations against network size for a selection of empirical networks spanning the whole range of sizes, and Table~\ref{tab:benchmark} summarises convergence rate, time and iterations of the three solvers for each model on 45 networks (Twitter crisis was not benchmarked), from $\sim10^2$ to $\sim4\times10^8$ dyads. In the benchmark a run stops when the maximum relative error is below $10^{-3}$ or after 5000 iterations. The coordinate solver converges for every model and network; fixed point is the fastest when it converges, while quasi-Newton is the least reliable for the \bipecm and the \abiecm (75\% and 84\% of the networks, respectively). Wall-clock time grows roughly linearly with the number of dyads for all solvers, and the \biecm and the \abiecm are one to two orders of magnitude slower than the \biwcm at fixed size. 

\input{SI_tabs/table_benchmark_v2}

\paragraph{Numerical uncertainty of the likelihood.} For a canonical model with constraints $\bm C$, parameters $\bm\theta$ and log-likelihood $\mathcal L(\bm\theta)=-\bm\theta^{\mathsf T}\bm C^*-\ln Z(\bm\theta)$, the score at an approximate solution is $\nabla\mathcal L=-[\bm C^*-\E_{\bm\theta}(\bm C)]$. A parameter perturbation $\delta\bm\theta$ therefore changes the likelihood to first order by $\delta\mathcal L\simeq-[\bm C^*-\E_{\bm\theta}(\bm C)]^{\mathsf T}\delta\bm\theta$. Close to the optimum the Hessian is minus the covariance matrix of the constraints, so the remaining likelihood gap can be estimated from the constraint residuals and the inverse covariance (with a pseudoinverse along the gauge directions). This estimate sets the scale below which likelihood or BIC differences should not be interpreted.

\end{document}

%% file: SI_tabs/table_systems_v2.tex
\begin{table}[t]
\centering
\scriptsize
\setlength{\tabcolsep}{3pt}
\begin{adjustbox}{max width=\textwidth}
\begin{tabular}{llllll}
\toprule
System & Layer 1 (rows, projected) & Layer 2 (columns) & Weight & $\Delta w$ & Snapshots \\
\midrule
Trade & Countries & Products & Export value & $10^{-12}$ & 2012--2020 (5) \\
Reddit & Subreddits & News domains & Times the domain is shared & 1 & 2013--2017 (5) \\
Lexical & Documents & Word stems & Stem occurrences & 1 & 2015--2023 (5) \\
Twitter & Verified users & Non-verified users & Retweets (counts) & 1 & 3 events \\
Tech CIT & Countries & Technology classes & Citations (counts) & $10^{-12}$ & 2010--2018 (5) \\
Tech CPC & Countries & CPC classes &  Fractional patent
counts & $10^{-12}$ & 2010--2018 (5) \\
Ecological & Plants / hosts & Animals / parasites & Visits, counts, intensity or prevalence & 1 or 0.1 & 21 matrices \\
\bottomrule
\end{tabular}
\end{adjustbox}
\caption{\textbf{Empirical systems.} Meaning of the two layers and of the weights, observation resolution $\Delta w$ used in the likelihoods (the largest power of ten such that every recorded weight is an integer multiple of it, capped at $10^{-12}$ for monetary and fractional data), and number of snapshots. Projections are always computed on layer~1. Sources and preprocessing are described in the text.}
\label{tab:systems}
\end{table}

%% file: SI_tabs/table_network_inputs_v2.tex
\begin{table}[p]
\centering
\scriptsize
\setlength{\tabcolsep}{3.5pt}
\renewcommand{\arraystretch}{1.02}
\begin{adjustbox}{max width=\textwidth}
\begin{tabular}{llrrrrrrrrr}
\toprule
Network & System & $N_1$ & $N_2$ & $L$ & $d$ & $\langle k\rangle$ & $\langle w\rangle$ & $\rho_{ks}$ & $Y$ & $\Delta w$ \\
\midrule
Trade 2012 & Trade & $220$ & $5204$ & $6.14\times10^{5}$ & $0.537$ & $227$ & $5.46\times10^{7}$ & $0.635$ & $0.187$ & $10^{-12}$ \\
Trade 2014 & Trade & $220$ & $5204$ & $5.26\times10^{5}$ & $0.46$ & $194$ & $3.52\times10^{7}$ & $0.595$ & $0.189$ & $10^{-12}$ \\
Trade 2016 & Trade & $220$ & $5204$ & $5.49\times10^{5}$ & $0.479$ & $202$ & $3.16\times10^{7}$ & $0.614$ & $0.191$ & $10^{-12}$ \\
Trade 2018 & Trade & $220$ & $5204$ & $5.61\times10^{5}$ & $0.49$ & $207$ & $3.76\times10^{7}$ & $0.597$ & $0.189$ & $10^{-12}$ \\
Trade 2020 & Trade & $220$ & $5204$ & $5.46\times10^{5}$ & $0.476$ & $201$ & $3.54\times10^{7}$ & $0.622$ & $0.19$ & $10^{-12}$ \\
\addlinespace[2pt]
Reddit 2013 & Reddit & $216$ & $6.88\times10^{4}$ & $1.59\times10^{5}$ & $0.0107$ & $4.61$ & $8.98$ & $0.861$ & $0.78$ & 1 \\
Reddit 2014 & Reddit & $248$ & $4.56\times10^{4}$ & $1.32\times10^{5}$ & $0.0116$ & $5.74$ & $8.62$ & $0.88$ & $0.758$ & 1 \\
Reddit 2015 & Reddit & $284$ & $4.54\times10^{4}$ & $1.48\times10^{5}$ & $0.0114$ & $6.46$ & $9.27$ & $0.886$ & $0.735$ & 1 \\
Reddit 2016 & Reddit & $399$ & $6.78\times10^{4}$ & $2.31\times10^{5}$ & $8.52\times10^{-3}$ & $6.76$ & $16.6$ & $0.868$ & $0.763$ & 1 \\
Reddit 2017 & Reddit & $461$ & $6.59\times10^{4}$ & $2.28\times10^{5}$ & $7.52\times10^{-3}$ & $6.88$ & $31.3$ & $0.854$ & $0.78$ & 1 \\
\addlinespace[2pt]
Lexical 2015 & Lexical & $40$ & $1.65\times10^{4}$ & $7.21\times10^{4}$ & $0.109$ & $8.73$ & $5.9$ & $0.866$ & $0.75$ & 1 \\
Lexical 2018 & Lexical & $56$ & $2.21\times10^{4}$ & $1.06\times10^{5}$ & $0.0858$ & $9.59$ & $6.16$ & $0.862$ & $0.768$ & 1 \\
Lexical 2020 & Lexical & $69$ & $2.74\times10^{4}$ & $1.43\times10^{5}$ & $0.0755$ & $10.4$ & $6.85$ & $0.853$ & $0.778$ & 1 \\
Lexical 2021 & Lexical & $76$ & $2.99\times10^{4}$ & $1.52\times10^{5}$ & $0.0669$ & $10.1$ & $6.82$ & $0.845$ & $0.788$ & 1 \\
Lexical 2023 & Lexical & $89$ & $3.54\times10^{4}$ & $1.83\times10^{5}$ & $0.0582$ & $10.3$ & $7.42$ & $0.839$ & $0.796$ & 1 \\
\addlinespace[2pt]
Twitter Q22 & Twitter & $1361$ & $6.56\times10^{4}$ & $1.45\times10^{5}$ & $1.62\times10^{-3}$ & $4.32$ & $1.44$ & $0.942$ & $0.767$ & 1 \\
Twitter election & Twitter & $2976$ & $1.32\times10^{5}$ & $4.60\times10^{5}$ & $1.17\times10^{-3}$ & $6.82$ & $1.99$ & $0.948$ & $0.688$ & 1 \\
Twitter crisis & Twitter & $6667$ & $3.22\times10^{5}$ & $8.85\times10^{5}$ & $4.13\times10^{-4}$ & $5.39$ & $1.47$ & $0.944$ & $0.745$ & 1 \\
\addlinespace[2pt]
Tech CIT 2010 & Tech CIT & $200$ & $312$ & $3.87\times10^{4}$ & $0.62$ & $151$ & $1795$ & $0.625$ & $0.111$ & $10^{-12}$ \\
Tech CIT 2012 & Tech CIT & $199$ & $312$ & $4.00\times10^{4}$ & $0.645$ & $157$ & $1441$ & $0.612$ & $0.104$ & $10^{-12}$ \\
Tech CIT 2014 & Tech CIT & $200$ & $312$ & $4.03\times10^{4}$ & $0.646$ & $158$ & $1041$ & $0.627$ & $0.0981$ & $10^{-12}$ \\
Tech CIT 2016 & Tech CIT & $204$ & $312$ & $4.00\times10^{4}$ & $0.629$ & $155$ & $539$ & $0.639$ & $0.0982$ & $10^{-12}$ \\
Tech CIT 2018 & Tech CIT & $190$ & $312$ & $3.29\times10^{4}$ & $0.555$ & $131$ & $67.6$ & $0.701$ & $0.11$ & $10^{-12}$ \\
\addlinespace[2pt]
Tech CPC 2010 & Tech CPC & $47$ & $650$ & $1.23\times10^{4}$ & $0.402$ & $35.3$ & $10.8$ & $0.882$ & $0.206$ & $10^{-12}$ \\
Tech CPC 2012 & Tech CPC & $47$ & $648$ & $1.30\times10^{4}$ & $0.426$ & $37.3$ & $10.7$ & $0.887$ & $0.202$ & $10^{-12}$ \\
Tech CPC 2014 & Tech CPC & $47$ & $648$ & $1.33\times10^{4}$ & $0.436$ & $38.2$ & $10.9$ & $0.89$ & $0.197$ & $10^{-12}$ \\
Tech CPC 2016 & Tech CPC & $47$ & $653$ & $1.35\times10^{4}$ & $0.439$ & $38.5$ & $10.5$ & $0.895$ & $0.193$ & $10^{-12}$ \\
Tech CPC 2018 & Tech CPC & $46$ & $620$ & $7216$ & $0.253$ & $21.7$ & $4.54$ & $0.903$ & $0.276$ & $10^{-12}$ \\
\addlinespace[2pt]
Beehler & Ecological & $11$ & $18$ & $38$ & $0.192$ & $2.62$ & $3.53$ & $0.755$ & $0.678$ & 1 \\
Seed disp. Beehler & Ecological & $9$ & $31$ & $119$ & $0.427$ & $5.95$ & $9.99$ & $0.902$ & $0.493$ & 1 \\
Seed disp. Snow & Ecological & $19$ & $29$ & $211$ & $0.383$ & $8.79$ & $94.5$ & $0.83$ & $0.397$ & 1 \\
Plant--pollinator BH & Ecological & $102$ & $12$ & $167$ & $0.136$ & $2.93$ & $3.29$ & $0.829$ & $0.786$ & 1 \\
Plants--ants & Ecological & $41$ & $51$ & $285$ & $0.136$ & $6.2$ & $2.26$ & $0.975$ & $0.424$ & 1 \\
Aishihik int. & Ecological & $7$ & $29$ & $78$ & $0.384$ & $4.33$ & $43.6$ & $0.74$ & $0.67$ & 0.1 \\
Aishihik prev. & Ecological & $7$ & $29$ & $78$ & $0.384$ & $4.33$ & $46$ & $0.837$ & $0.592$ & 0.1 \\
Canada & Ecological & $10$ & $38$ & $87$ & $0.229$ & $3.62$ & $18.9$ & $0.753$ & $0.673$ & 1 \\
Cold Lake int. & Ecological & $10$ & $40$ & $91$ & $0.228$ & $3.64$ & $18.4$ & $0.744$ & $0.678$ & 0.1 \\
Cold Lake prev. & Ecological & $10$ & $40$ & $91$ & $0.228$ & $3.64$ & $24.9$ & $0.763$ & $0.719$ & 0.1 \\
Huron prev. & Ecological & $33$ & $97$ & $316$ & $0.0987$ & $4.86$ & $25.5$ & $0.79$ & $0.585$ & 0.1 \\
Lake of the Woods & Ecological & $31$ & $146$ & $446$ & $0.0985$ & $5.04$ & $29.7$ & $0.836$ & $0.615$ & 0.1 \\
Lake of the Woods prev. & Ecological & $31$ & $144$ & $384$ & $0.086$ & $4.39$ & $31.6$ & $0.824$ & $0.627$ & 0.1 \\
McGregor$^{a}$ & Ecological & $14$ & $51$ & $114$ & $0.16$ & $3.51$ & $11.5$ & $0.774$ & $0.701$ & 0.1 \\
Parsnip int. & Ecological & $17$ & $53$ & $158$ & $0.175$ & $4.51$ & $12.7$ & $0.776$ & $0.578$ & 0.1 \\
Parsnip prev. & Ecological & $17$ & $53$ & $158$ & $0.175$ & $4.51$ & $9.84$ & $0.781$ & $0.586$ & 0.1 \\
Smallwood int. & Ecological & $6$ & $25$ & $53$ & $0.353$ & $3.42$ & $17.8$ & $0.697$ & $0.687$ & 1 \\
Smallwood prev. & Ecological & $6$ & $25$ & $53$ & $0.353$ & $3.42$ & $24.4$ & $0.683$ & $0.639$ & 1 \\
\bottomrule
\end{tabular}
\end{adjustbox}
\caption{\textbf{Basic statistics of the empirical bipartite networks.} Number of nodes in the two layers ($N_1,N_2$), number of links $L$, density $d=L/(N_1N_2)$, average degree $\langle k\rangle=2L/(N_1+N_2)$, average positive weight $\langle w\rangle$, Spearman correlation between strengths and degrees $\rho_{ks}$, weight disparity $Y$ (Methods) and observation resolution $\Delta w$. $^{a}$~Intensity matrix; the prevalence matrix in the source data is identical to it and is not included.}
\label{tab:input_network_statistics}
\end{table}

%% file: SI_tabs/table_reconstruction_representative_v2.tex
\begin{table}[t]
\centering
\scriptsize
\setlength{\tabcolsep}{3pt}
\begin{adjustbox}{max width=\textwidth}
\begin{tabular}{llrrrrrr}
\toprule
& & \multicolumn{3}{c}{$k^{nn}$} & \multicolumn{3}{c}{$s^{nn}$} \\
\cmidrule(lr){3-5}\cmidrule(lr){6-8}
Network & Model & $\mathrm{RMSE}_z$ & med.\ rel.\ err. & in 95\% & $\mathrm{RMSE}_z$ & med.\ rel.\ err. & in 95\% \\
\midrule
Trade 2020 & BiWCM & -- & -- & -- & $48.48$ & $0.40$ & $0.20$ \\
 & P-BiECM & $57.99$ & $0.17$ & $4.5\times10^{-3}$ & $10.50$ & $0.31$ & $0.13$ \\
 & BiECM & $4.37$ & $7.7\times10^{-3}$ & $0.49$ & $0.87$ & $0.03$ & $0.96$ \\
\addlinespace[2pt]
Lake of the Woods & BiWCM & -- & -- & -- & $70.50$ & $0.69$ & $0.00$ \\
 & P-BiECM & $10.15$ & $0.58$ & $0.03$ & $2.30$ & $0.41$ & $0.58$ \\
 & BiECM & $0.99$ & $0.10$ & $0.87$ & $0.87$ & $0.08$ & $0.90$ \\
\midrule
& & \multicolumn{2}{c}{NODF} & \multicolumn{2}{c}{$C$} & \multicolumn{2}{c}{WNODA} \\
\cmidrule(lr){3-4}\cmidrule(lr){5-6}\cmidrule(lr){7-8}
Network & Model & $z$ & perc. & $z$ & perc. & $z$ & perc. \\
\midrule
Trade 2020 & BiWCM & -- & -- & -- & -- & $-161.3$ & $0.0$ \\
 & P-BiECM & $367.6$ & $100.0$ & $172.6$ & $100.0$ & $209.9$ & $100.0$ \\
 & BiECM & $-15.1$ & $0.0$ & $6.4$ & $100.0$ & $-53.7$ & $0.0$ \\
\addlinespace[2pt]
Lake of the Woods & BiWCM & -- & -- & -- & -- & $-183.7$ & $0.0$ \\
 & P-BiECM & $22.2$ & $100.0$ & $16.6$ & $100.0$ & $3.7$ & $100.0$ \\
 & BiECM & $-1.1$ & $14.4$ & $-0.4$ & $35.2$ & $1.8$ & $96.5$ \\
\bottomrule
\end{tabular}
\end{adjustbox}
\caption{\textbf{Reconstruction of the two networks of the main-text figure.} $2000$ networks sampled from each fitted model. Top: nearest-neighbour degree and strength of the row nodes, with $\mathrm{RMSE}_z$, the median relative error $|x^*_i-\langle x_i\rangle|/x^*_i$ and the fraction of nodes whose observed value falls in the central 95\% interval of the sampled values. Bottom: z-score $(y^*-\langle y\rangle)/\sigma_y$ and percentile of the observed value in the sampled distribution for NODF, Robins--Alexander clustering $C$ and WNODA. BiWCM networks are complete, so $k^{nn}$, NODF and $C$ are not random under that model (--). The weighted clustering $C_W$ is not shown because it equals one in Trade 2020 under every model; in Lake of the Woods its z-score is $5.0$ for the P-BiECM and $-0.39$ for the BiECM.}
\label{tab:reconstruction_representative}
\end{table}

%% file: SI_tabs/table_model_selection_v3.tex
\begin{table}[p]
\centering
\scriptsize
\setlength{\tabcolsep}{2.6pt}
\renewcommand{\arraystretch}{1.02}
\begin{adjustbox}{max width=\textwidth}
\begin{tabular}{lrrrrrrrcc}
\toprule
& \multicolumn{3}{c}{$\ln\mathcal{L}$} & \multicolumn{3}{c}{BIC} & & \multicolumn{2}{c}{Selected} \\
\cmidrule(lr){2-4}\cmidrule(lr){5-7}\cmidrule(lr){9-10}
Network & BiWCM & P-BiECM & BiECM & BiWCM & P-BiECM & BiECM & $\Delta\mathrm{BIC}_{2}$ & BIC & AIC \\
\midrule
Trade 2012 & $-4.86\times10^{7}$ & $-2.78\times10^{7}$ & {\bfseries\boldmath $-2.75\times10^{7}$} & $9.72\times10^{7}$ & $5.57\times10^{7}$ & {\bfseries\boldmath $5.52\times10^{7}$} & $4.41\times10^{5}$ & BiECM & BiECM \\
Trade 2014 & $-4.79\times10^{7}$ & $-2.38\times10^{7}$ & {\bfseries\boldmath $-2.35\times10^{7}$} & $9.58\times10^{7}$ & $4.76\times10^{7}$ & {\bfseries\boldmath $4.71\times10^{7}$} & $4.60\times10^{5}$ & BiECM & BiECM \\
Trade 2016 & $-4.78\times10^{7}$ & $-2.47\times10^{7}$ & {\bfseries\boldmath $-2.44\times10^{7}$} & $9.57\times10^{7}$ & $4.94\times10^{7}$ & {\bfseries\boldmath $4.90\times10^{7}$} & $4.33\times10^{5}$ & BiECM & BiECM \\
Trade 2018 & $-4.80\times10^{7}$ & $-2.53\times10^{7}$ & {\bfseries\boldmath $-2.50\times10^{7}$} & $9.61\times10^{7}$ & $5.06\times10^{7}$ & {\bfseries\boldmath $5.02\times10^{7}$} & $4.52\times10^{5}$ & BiECM & BiECM \\
Trade 2020 & $-4.79\times10^{7}$ & $-2.46\times10^{7}$ & {\bfseries\boldmath $-2.43\times10^{7}$} & $9.59\times10^{7}$ & $4.92\times10^{7}$ & {\bfseries\boldmath $4.87\times10^{7}$} & $4.50\times10^{5}$ & BiECM & BiECM \\
\addlinespace[2pt]
Reddit 2013 & $-1.50\times10^{7}$ & $-9.50\times10^{5}$ & {\bfseries\boldmath $-6.72\times10^{5}$} & $3.11\times10^{7}$ & {\bfseries\boldmath $3.04\times10^{6}$} & $3.62\times10^{6}$ & $5.84\times10^{5}$ & P-BiECM & BiECM \\
Reddit 2014 & $-1.14\times10^{7}$ & $-7.89\times10^{5}$ & {\bfseries\boldmath $-6.32\times10^{5}$} & $2.36\times10^{7}$ & {\bfseries\boldmath $2.32\times10^{6}$} & $2.75\times10^{6}$ & $4.30\times10^{5}$ & P-BiECM & BiECM \\
Reddit 2015 & $-1.30\times10^{7}$ & $-8.97\times10^{5}$ & {\bfseries\boldmath $-7.42\times10^{5}$} & $2.68\times10^{7}$ & {\bfseries\boldmath $2.54\times10^{6}$} & $2.98\times10^{6}$ & $4.37\times10^{5}$ & P-BiECM & BiECM \\
Reddit 2016 & $-2.73\times10^{7}$ & $-1.48\times10^{6}$ & {\bfseries\boldmath $-1.20\times10^{6}$} & $5.58\times10^{7}$ & {\bfseries\boldmath $4.13\times10^{6}$} & $4.73\times10^{6}$ & $5.95\times10^{5}$ & P-BiECM & BiECM \\
Reddit 2017 & $-3.07\times10^{7}$ & $-1.53\times10^{6}$ & {\bfseries\boldmath $-1.22\times10^{6}$} & $6.25\times10^{7}$ & {\bfseries\boldmath $4.19\times10^{6}$} & $4.72\times10^{6}$ & $5.29\times10^{5}$ & P-BiECM & BiECM \\
\addlinespace[2pt]
Lexical 2015 & $-7.49\times10^{5}$ & $-3.11\times10^{5}$ & {\bfseries\boldmath $-2.81\times10^{5}$} & $1.72\times10^{6}$ & {\bfseries\boldmath $8.42\times10^{5}$} & $1.00\times10^{6}$ & $1.63\times10^{5}$ & P-BiECM & BiECM \\
Lexical 2018 & $-1.37\times10^{6}$ & $-4.69\times10^{5}$ & {\bfseries\boldmath $-4.22\times10^{5}$} & $3.05\times10^{6}$ & {\bfseries\boldmath $1.25\times10^{6}$} & $1.47\times10^{6}$ & $2.17\times10^{5}$ & P-BiECM & BiECM \\
Lexical 2020 & $-2.08\times10^{6}$ & $-6.45\times10^{5}$ & {\bfseries\boldmath $-5.78\times10^{5}$} & $4.56\times10^{6}$ & {\bfseries\boldmath $1.69\times10^{6}$} & $1.95\times10^{6}$ & $2.62\times10^{5}$ & P-BiECM & BiECM \\
Lexical 2021 & $-2.47\times10^{6}$ & $-6.97\times10^{5}$ & {\bfseries\boldmath $-6.20\times10^{5}$} & $5.37\times10^{6}$ & {\bfseries\boldmath $1.83\times10^{6}$} & $2.12\times10^{6}$ & $2.85\times10^{5}$ & P-BiECM & BiECM \\
Lexical 2023 & $-3.39\times10^{6}$ & $-8.65\times10^{5}$ & {\bfseries\boldmath $-7.66\times10^{5}$} & $7.31\times10^{6}$ & {\bfseries\boldmath $2.26\times10^{6}$} & $2.59\times10^{6}$ & $3.32\times10^{5}$ & P-BiECM & BiECM \\
\addlinespace[2pt]
Twitter Q22 & $-8.93\times10^{7}$ & $-9.87\times10^{5}$ & {\bfseries\boldmath $-8.42\times10^{5}$} & $1.80\times10^{8}$ & {\bfseries\boldmath $3.20\times10^{6}$} & $4.14\times10^{6}$ & $9.36\times10^{5}$ & P-BiECM & BiECM \\
Twitter election & $-3.92\times10^{8}$ & $-3.05\times10^{6}$ & {\bfseries\boldmath $-2.53\times10^{6}$} & $7.87\times10^{8}$ & {\bfseries\boldmath $8.76\times10^{6}$} & $1.04\times10^{7}$ & $1.64\times10^{6}$ & P-BiECM & BiECM \\
Twitter crisis & $-2.15\times10^{9}$ & $-6.97\times10^{6}$ & {\bfseries\boldmath $-5.87\times10^{6}$} & $4.30\times10^{9}$ & {\bfseries\boldmath $2.10\times10^{7}$} & $2.59\times10^{7}$ & $4.87\times10^{6}$ & P-BiECM & BiECM \\
\addlinespace[2pt]
Tech CIT 2010 & $-1.94\times10^{6}$ & $-1.30\times10^{6}$ & {\bfseries\boldmath $-1.29\times10^{6}$} & $3.88\times10^{6}$ & $2.61\times10^{6}$ & {\bfseries\boldmath $2.60\times10^{6}$} & $1.06\times10^{4}$ & BiECM & BiECM \\
Tech CIT 2012 & $-1.92\times10^{6}$ & $-1.34\times10^{6}$ & {\bfseries\boldmath $-1.33\times10^{6}$} & $3.84\times10^{6}$ & $2.68\times10^{6}$ & {\bfseries\boldmath $2.67\times10^{6}$} & $1.07\times10^{4}$ & BiECM & BiECM \\
Tech CIT 2014 & $-1.91\times10^{6}$ & $-1.34\times10^{6}$ & {\bfseries\boldmath $-1.33\times10^{6}$} & $3.83\times10^{6}$ & $2.68\times10^{6}$ & {\bfseries\boldmath $2.67\times10^{6}$} & $1.04\times10^{4}$ & BiECM & BiECM \\
Tech CIT 2016 & $-1.91\times10^{6}$ & $-1.30\times10^{6}$ & {\bfseries\boldmath $-1.29\times10^{6}$} & $3.82\times10^{6}$ & $2.61\times10^{6}$ & {\bfseries\boldmath $2.60\times10^{6}$} & $1.14\times10^{4}$ & BiECM & BiECM \\
Tech CIT 2018 & $-1.66\times10^{6}$ & $-1.02\times10^{6}$ & {\bfseries\boldmath $-1.01\times10^{6}$} & $3.33\times10^{6}$ & $2.05\times10^{6}$ & {\bfseries\boldmath $2.04\times10^{6}$} & $1.10\times10^{4}$ & BiECM & BiECM \\
\addlinespace[2pt]
Tech CPC 2010 & $-8.39\times10^{5}$ & $-3.80\times10^{5}$ & {\bfseries\boldmath $-3.76\times10^{5}$} & $1.69\times10^{6}$ & $7.66\times10^{5}$ & {\bfseries\boldmath $7.66\times10^{5}$} & $591$ & BiECM & BiECM \\
Tech CPC 2012 & $-8.38\times10^{5}$ & $-3.99\times10^{5}$ & {\bfseries\boldmath $-3.95\times10^{5}$} & $1.68\times10^{6}$ & $8.06\times10^{5}$ & {\bfseries\boldmath $8.05\times10^{5}$} & $674$ & BiECM & BiECM \\
Tech CPC 2014 & $-8.40\times10^{5}$ & $-4.08\times10^{5}$ & {\bfseries\boldmath $-4.04\times10^{5}$} & $1.69\times10^{6}$ & $8.23\times10^{5}$ & {\bfseries\boldmath $8.23\times10^{5}$} & $786$ & BiECM & BiECM \\
Tech CPC 2016 & $-8.47\times10^{5}$ & $-4.14\times10^{5}$ & {\bfseries\boldmath $-4.10\times10^{5}$} & $1.70\times10^{6}$ & $8.36\times10^{5}$ & {\bfseries\boldmath $8.35\times10^{5}$} & $854$ & BiECM & BiECM \\
Tech CPC 2018 & $-7.41\times10^{5}$ & $-2.21\times10^{5}$ & {\bfseries\boldmath $-2.19\times10^{5}$} & $1.49\times10^{6}$ & {\bfseries\boldmath $4.49\times10^{5}$} & $4.51\times10^{5}$ & $1373$ & P-BiECM & BiECM \\
\addlinespace[2pt]
Beehler & $-236$ & $-150$ & {\bfseries\boldmath $-137$} & $626$ & {\bfseries\boldmath $458$} & $581$ & $123$ & P-BiECM & P-BiECM \\
Seed disp. Beehler & $-501$ & $-450$ & {\bfseries\boldmath $-420$} & $1227$ & {\bfseries\boldmath $1131$} & $1291$ & $95.9$ & P-BiECM & P-BiECM \\
Seed disp. Snow & $-1464$ & $-1199$ & {\bfseries\boldmath $-1112$} & $3230$ & {\bfseries\boldmath $2706$} & $2830$ & $123$ & P-BiECM & BiECM \\
Plant--pollinator BH & $-1351$ & $-675$ & {\bfseries\boldmath $-557$} & $3513$ & {\bfseries\boldmath $2167$} & $2734$ & $568$ & P-BiECM & BiECM \\
Plants--ants & $-2256$ & $-1039$ & {\bfseries\boldmath $-928$} & $5216$ & {\bfseries\boldmath $2788$} & $3262$ & $474$ & P-BiECM & BiECM \\
Aishihik int. & $-802$ & $-559$ & {\bfseries\boldmath $-517$} & $1795$ & {\bfseries\boldmath $1315$} & $1416$ & $101$ & P-BiECM & BiECM \\
Aishihik prev. & $-1136$ & $-665$ & {\bfseries\boldmath $-623$} & $2463$ & {\bfseries\boldmath $1526$} & $1629$ & $104$ & P-BiECM & BiECM \\
Canada & $-602$ & $-433$ & {\bfseries\boldmath $-398$} & $1489$ & {\bfseries\boldmath $1158$} & $1366$ & $209$ & P-BiECM & P-BiECM \\
Cold Lake int. & $-1279$ & $-667$ & {\bfseries\boldmath $-624$} & $2858$ & {\bfseries\boldmath $1639$} & $1846$ & $208$ & P-BiECM & P-BiECM \\
Cold Lake prev. & $-1674$ & $-761$ & {\bfseries\boldmath $-712$} & $3648$ & {\bfseries\boldmath $1828$} & $2024$ & $195$ & P-BiECM & BiECM \\
Huron prev. & $-1.03\times10^{4}$ & $-2903$ & {\bfseries\boldmath $-2689$} & $2.17\times10^{4}$ & {\bfseries\boldmath $6864$} & $7476$ & $613$ & P-BiECM & BiECM \\
Lake of the Woods & $-1.67\times10^{4}$ & $-4252$ & {\bfseries\boldmath $-3912$} & $3.49\times10^{4}$ & {\bfseries\boldmath $1.00\times10^{4}$} & $1.08\times10^{4}$ & $802$ & P-BiECM & BiECM \\
Lake of the Woods prev. & $-1.62\times10^{4}$ & $-3745$ & {\bfseries\boldmath $-3522$} & $3.39\times10^{4}$ & {\bfseries\boldmath $8968$} & $9986$ & $1018$ & P-BiECM & BiECM \\
McGregor$^{a}$ & $-1926$ & $-864$ & {\bfseries\boldmath $-801$} & $4279$ & {\bfseries\boldmath $2163$} & $2456$ & $293$ & P-BiECM & P-BiECM \\
Parsnip int. & $-3089$ & $-1273$ & {\bfseries\boldmath $-1184$} & $6655$ & {\bfseries\boldmath $3029$} & $3320$ & $290$ & P-BiECM & BiECM \\
Parsnip prev. & $-2602$ & $-1190$ & {\bfseries\boldmath $-1119$} & $5680$ & {\bfseries\boldmath $2864$} & $3191$ & $327$ & P-BiECM & BiECM \\
Smallwood int. & $-339$ & $-269$ & {\bfseries\boldmath $-238$} & $834$ & {\bfseries\boldmath $699$} & $787$ & $88.3$ & P-BiECM & BiECM \\
Smallwood prev. & $-414$ & $-301$ & {\bfseries\boldmath $-272$} & $983$ & {\bfseries\boldmath $762$} & $855$ & $93.7$ & P-BiECM & P-BiECM \\
\bottomrule
\end{tabular}
\end{adjustbox}
\caption{\textbf{Likelihood-based model comparison.} Log-likelihood of the recorded data, evaluated at the fitted parameters, with each empty cell given the probability $1-p_{i\alpha}e^{-\lambda_{i\alpha}\Delta w}$ of a value below $\Delta w$ (main text, Methods) (probabilities integrated over the observation bins of width $\Delta w$, Table~\ref{tab:input_network_statistics}) and $\mathrm{BIC}=k\ln(N_1N_2)-2\ln\mathcal{L}$, with $k=N_1+N_2$ (BiWCM), $N_1+N_2+1$ (P-BiECM) and $2(N_1+N_2)$ (BiECM). The largest likelihood and the smallest BIC are in bold. $\Delta\mathrm{BIC}_{2}$ is the gap between the best and the second-best model: since its smallest value over the corpus is $88.3$, the BIC weight of the selected model, $[1+\sum_{m\neq\mathrm{best}}e^{-\Delta\mathrm{BIC}_m/2}]^{-1}$, exceeds $1-10^{-8}$ in every network. The last two columns give the model selected by BIC and by AIC$=2k-2\ln\mathcal{L}$. $^{a}$~Intensity matrix; the prevalence matrix in the source data is identical to it and is not included.}
\label{tab:model_selection}
\end{table}

%% file: SI_tabs/table_reconstruction_errors_v2.tex
\begin{table}[p]
\centering
\scriptsize
\setlength{\tabcolsep}{2.4pt}
\renewcommand{\arraystretch}{1.02}
\begin{adjustbox}{max width=\textwidth}
\begin{tabular}{lrrrrrrrrrrrr}
\toprule
 & \multicolumn{3}{c}{$k^{nn}$, layer 1} & \multicolumn{3}{c}{$k^{nn}$, layer 2} & \multicolumn{3}{c}{$s^{nn}$, layer 1} & \multicolumn{3}{c}{$s^{nn}$, layer 2} \\
\cmidrule(lr){2-4}\cmidrule(lr){5-7}\cmidrule(lr){8-10}\cmidrule(lr){11-13}
Network & BiWCM & P-BiECM & BiECM & BiWCM & P-BiECM & BiECM & BiWCM & P-BiECM & BiECM & BiWCM & P-BiECM & BiECM \\
\midrule
Trade 2012 & -- & $63.5$ & {\bfseries\boldmath $3.61$} & -- & $9.26$ & {\bfseries\boldmath $1.38$} & $25.5$ & $7.83$ & {\bfseries\boldmath $0.743$} & $21.9$ & $4.61$ & {\bfseries\boldmath $0.336$} \\
Trade 2014 & -- & $59.8$ & {\bfseries\boldmath $3.92$} & -- & $10.8$ & {\bfseries\boldmath $1.51$} & $32.5$ & $8.1$ & {\bfseries\boldmath $0.861$} & $27.9$ & $4.53$ & {\bfseries\boldmath $0.398$} \\
Trade 2016 & -- & $61.9$ & {\bfseries\boldmath $4.12$} & -- & $10.2$ & {\bfseries\boldmath $1.44$} & $46.6$ & $10$ & {\bfseries\boldmath $0.886$} & $39.8$ & $4.93$ & {\bfseries\boldmath $0.433$} \\
Trade 2018 & -- & $61.3$ & {\bfseries\boldmath $4.41$} & -- & $9.98$ & {\bfseries\boldmath $1.46$} & $37.8$ & $8.27$ & {\bfseries\boldmath $0.86$} & $34.2$ & $4.91$ & {\bfseries\boldmath $0.41$} \\
Trade 2020 & -- & $58.5$ & {\bfseries\boldmath $4.38$} & -- & $10.8$ & {\bfseries\boldmath $1.44$} & $46.9$ & $10.4$ & {\bfseries\boldmath $0.942$} & $40.8$ & $5.13$ & {\bfseries\boldmath $0.458$} \\
\addlinespace[2pt]
Reddit 2013 & -- & $274$ & {\bfseries\boldmath $6.07$} & -- & $52.6$ & {\bfseries\boldmath $0.859$} & $6296$ & $239$ & {\bfseries\boldmath $0.686$} & $821$ & $11.3$ & {\bfseries\boldmath $0.84$} \\
Reddit 2014 & -- & $187$ & {\bfseries\boldmath $5.21$} & -- & $30.1$ & {\bfseries\boldmath $0.991$} & $4217$ & $115$ & {\bfseries\boldmath $1.06$} & $446$ & $6.91$ & {\bfseries\boldmath $1$} \\
Reddit 2015 & -- & $188$ & {\bfseries\boldmath $5.32$} & -- & $20$ & {\bfseries\boldmath $0.955$} & $4050$ & $95.2$ & {\bfseries\boldmath $0.954$} & $191$ & $3.56$ & {\bfseries\boldmath $1.03$} \\
Reddit 2016 & -- & $232$ & {\bfseries\boldmath $5.8$} & -- & $39.1$ & {\bfseries\boldmath $1.2$} & $7762$ & $195$ & {\bfseries\boldmath $1.11$} & $521$ & $12.6$ & {\bfseries\boldmath $1.21$} \\
Reddit 2017 & -- & $239$ & {\bfseries\boldmath $5.38$} & -- & $49.8$ & {\bfseries\boldmath $1.18$} & $6479$ & $140$ & {\bfseries\boldmath $1.03$} & $476$ & $9.07$ & {\bfseries\boldmath $1.22$} \\
\addlinespace[2pt]
Lexical 2015 & -- & $70.7$ & {\bfseries\boldmath $3.69$} & -- & $3.28$ & {\bfseries\boldmath $0.895$} & $527$ & $18.8$ & {\bfseries\boldmath $0.902$} & $55.8$ & $1.63$ & {\bfseries\boldmath $0.991$} \\
Lexical 2018 & -- & $82.6$ & {\bfseries\boldmath $3.95$} & -- & $3$ & {\bfseries\boldmath $0.859$} & $1061$ & $25$ & {\bfseries\boldmath $1.26$} & $74$ & $1.63$ & {\bfseries\boldmath $0.987$} \\
Lexical 2020 & -- & $90.4$ & {\bfseries\boldmath $3.94$} & -- & $3.34$ & {\bfseries\boldmath $0.854$} & $1136$ & $23.4$ & {\bfseries\boldmath $1.07$} & $72.9$ & $1.74$ & {\bfseries\boldmath $0.997$} \\
Lexical 2021 & -- & $98.8$ & {\bfseries\boldmath $3.66$} & -- & $3.48$ & {\bfseries\boldmath $0.913$} & $1452$ & $29.6$ & {\bfseries\boldmath $1.1$} & $107$ & $2.07$ & {\bfseries\boldmath $1.09$} \\
Lexical 2023 & -- & $106$ & {\bfseries\boldmath $4$} & -- & $4.26$ & {\bfseries\boldmath $0.897$} & $1438$ & $29.3$ & {\bfseries\boldmath $1.05$} & $113$ & $2.35$ & {\bfseries\boldmath $1.11$} \\
\addlinespace[2pt]
Twitter Q22 & -- & $134$ & {\bfseries\boldmath $5.07$} & -- & $16.9$ & {\bfseries\boldmath $1.31$} & $3844$ & $31.8$ & {\bfseries\boldmath $3.17$} & $5628$ & $4.43$ & {\bfseries\boldmath $1.27$} \\
Twitter election & -- & $64.3$ & {\bfseries\boldmath $3.91$} & -- & $16.6$ & {\bfseries\boldmath $1.25$} & $4199$ & $26.7$ & {\bfseries\boldmath $3.06$} & $1.71\times10^{4}$ & $4.57$ & {\bfseries\boldmath $1.23$} \\
Twitter crisis & -- & $69.3$ & {\bfseries\boldmath $4.53$} & -- & $30.1$ & {\bfseries\boldmath $1.42$} & $3403$ & $17.3$ & {\bfseries\boldmath $3.52$} & $4.44\times10^{4}$ & $10.3$ & {\bfseries\boldmath $1.4$} \\
\addlinespace[2pt]
Tech CIT 2010 & -- & $19.1$ & {\bfseries\boldmath $1.18$} & -- & $5.34$ & {\bfseries\boldmath $0.822$} & $13.7$ & $3.59$ & {\bfseries\boldmath $0.565$} & $13.3$ & $2.7$ & {\bfseries\boldmath $0.0672$} \\
Tech CIT 2012 & -- & $18.7$ & {\bfseries\boldmath $1.13$} & -- & $5.26$ & {\bfseries\boldmath $0.799$} & $14.7$ & $3.64$ & {\bfseries\boldmath $0.555$} & $12.2$ & $2.64$ & {\bfseries\boldmath $0.105$} \\
Tech CIT 2014 & -- & $18.7$ & {\bfseries\boldmath $1.22$} & -- & $5.19$ & {\bfseries\boldmath $0.831$} & $15$ & $3.74$ & {\bfseries\boldmath $0.569$} & $13$ & $2.74$ & {\bfseries\boldmath $0.0935$} \\
Tech CIT 2016 & -- & $18.3$ & {\bfseries\boldmath $1.13$} & -- & $5.37$ & {\bfseries\boldmath $0.882$} & $14.7$ & $4.01$ & {\bfseries\boldmath $0.529$} & $15.3$ & $2.97$ & {\bfseries\boldmath $0.0915$} \\
Tech CIT 2018 & -- & $18.1$ & {\bfseries\boldmath $1.24$} & -- & $5.47$ & {\bfseries\boldmath $0.886$} & $18.2$ & $4.15$ & {\bfseries\boldmath $0.576$} & $24.5$ & $3.66$ & {\bfseries\boldmath $0.16$} \\
\addlinespace[2pt]
Tech CPC 2010 & -- & $15$ & {\bfseries\boldmath $1.55$} & -- & $4.57$ & {\bfseries\boldmath $0.716$} & $38$ & $7.31$ & {\bfseries\boldmath $0.621$} & $37.5$ & $2.61$ & {\bfseries\boldmath $0.421$} \\
Tech CPC 2012 & -- & $14.1$ & {\bfseries\boldmath $1.43$} & -- & $4.44$ & {\bfseries\boldmath $0.758$} & $31.6$ & $6.2$ & {\bfseries\boldmath $0.518$} & $38.2$ & $2.53$ & {\bfseries\boldmath $0.413$} \\
Tech CPC 2014 & -- & $14.5$ & {\bfseries\boldmath $1.44$} & -- & $4.55$ & {\bfseries\boldmath $0.752$} & $26.6$ & $6.13$ & {\bfseries\boldmath $0.345$} & $30.7$ & $2.49$ & {\bfseries\boldmath $0.376$} \\
Tech CPC 2016 & -- & $14.7$ & {\bfseries\boldmath $1.38$} & -- & $4.59$ & {\bfseries\boldmath $0.73$} & $33.2$ & $6.92$ & {\bfseries\boldmath $0.377$} & $36$ & $2.6$ & {\bfseries\boldmath $0.398$} \\
Tech CPC 2018 & -- & $11.3$ & {\bfseries\boldmath $1.48$} & -- & $4.47$ & {\bfseries\boldmath $0.758$} & $51.6$ & $6.56$ & {\bfseries\boldmath $0.803$} & $64.7$ & $2.39$ & {\bfseries\boldmath $0.65$} \\
\addlinespace[2pt]
Beehler & -- & $1.5$ & {\bfseries\boldmath $0.921$} & -- & $1.13$ & {\bfseries\boldmath $0.625$} & $5.28$ & $0.666$ & {\bfseries\boldmath $0.481$} & $5.86$ & $0.716$ & {\bfseries\boldmath $0.558$} \\
Seed disp. Beehler & -- & $1.85$ & {\bfseries\boldmath $1.01$} & -- & $1.42$ & {\bfseries\boldmath $0.755$} & $5.76$ & $0.988$ & {\bfseries\boldmath $0.505$} & $4.81$ & $1.14$ & {\bfseries\boldmath $0.676$} \\
Seed disp. Snow & -- & $1.47$ & {\bfseries\boldmath $0.701$} & -- & $2.36$ & {\bfseries\boldmath $0.627$} & $5.31$ & $1.34$ & {\bfseries\boldmath $0.811$} & $7.07$ & $0.798$ & {\bfseries\boldmath $0.494$} \\
Plant--pollinator BH & -- & $4.24$ & {\bfseries\boldmath $0.931$} & -- & $3.49$ & {\bfseries\boldmath $1.61$} & $12.4$ & $1.61$ & {\bfseries\boldmath $0.833$} & $30.4$ & $3.48$ & {\bfseries\boldmath $0.617$} \\
Plants--ants & -- & $4.63$ & {\bfseries\boldmath $0.952$} & -- & $3.63$ & {\bfseries\boldmath $0.846$} & $32.3$ & $1.92$ & {\bfseries\boldmath $0.718$} & $27.1$ & $2.1$ & {\bfseries\boldmath $0.844$} \\
Aishihik int. & -- & {\bfseries\boldmath $0.689$} & $1.83$ & -- & $0.972$ & {\bfseries\boldmath $0.525$} & $4.85$ & $0.827$ & {\bfseries\boldmath $0.265$} & $2.46$ & $0.956$ & {\bfseries\boldmath $0.955$} \\
Aishihik prev. & -- & {\bfseries\boldmath $1.23$} & $2.06$ & -- & $1.2$ & {\bfseries\boldmath $0.511$} & $5.77$ & {\bfseries\boldmath $0.516$} & $0.556$ & $2.63$ & $0.749$ & {\bfseries\boldmath $0.437$} \\
Canada & -- & {\bfseries\boldmath $0.722$} & $1.39$ & -- & $1.14$ & {\bfseries\boldmath $0.645$} & $9.64$ & $1.14$ & {\bfseries\boldmath $0.229$} & $1.67$ & $0.551$ & {\bfseries\boldmath $0.548$} \\
Cold Lake int. & -- & {\bfseries\boldmath $0.779$} & $1.39$ & -- & $1.25$ & {\bfseries\boldmath $0.652$} & $11.6$ & $1.13$ & {\bfseries\boldmath $0.209$} & $1.95$ & {\bfseries\boldmath $0.561$} & $0.574$ \\
Cold Lake prev. & -- & {\bfseries\boldmath $0.762$} & $1.35$ & -- & $1.47$ & {\bfseries\boldmath $0.697$} & $11.6$ & $1.04$ & {\bfseries\boldmath $0.463$} & $2.09$ & $0.667$ & {\bfseries\boldmath $0.592$} \\
Huron prev. & -- & $5.44$ & {\bfseries\boldmath $1.07$} & -- & $2.57$ & {\bfseries\boldmath $0.895$} & $93.5$ & $3.17$ & {\bfseries\boldmath $0.825$} & $8.57$ & $1.01$ & {\bfseries\boldmath $0.721$} \\
Lake of the Woods & -- & $9.97$ & {\bfseries\boldmath $0.993$} & -- & $1.56$ & {\bfseries\boldmath $0.859$} & $68.1$ & $2.3$ & {\bfseries\boldmath $0.845$} & $12.6$ & $0.898$ & {\bfseries\boldmath $0.735$} \\
Lake of the Woods prev. & -- & $5.37$ & {\bfseries\boldmath $1.24$} & -- & $1.6$ & {\bfseries\boldmath $0.879$} & $56.6$ & $1.75$ & {\bfseries\boldmath $0.874$} & $13.3$ & $0.858$ & {\bfseries\boldmath $0.742$} \\
McGregor$^{a}$ & -- & $1.74$ & {\bfseries\boldmath $1.23$} & -- & $1.75$ & {\bfseries\boldmath $0.644$} & $18.8$ & $1.32$ & {\bfseries\boldmath $0.55$} & $3.93$ & $0.802$ & {\bfseries\boldmath $0.65$} \\
Parsnip int. & -- & $2.86$ & {\bfseries\boldmath $1.23$} & -- & $1.05$ & {\bfseries\boldmath $0.602$} & $24.6$ & $2.03$ & {\bfseries\boldmath $1.02$} & $5.48$ & $0.713$ & {\bfseries\boldmath $0.615$} \\
Parsnip prev. & -- & $2.14$ & {\bfseries\boldmath $1.27$} & -- & $1.07$ & {\bfseries\boldmath $0.642$} & $15.7$ & $0.906$ & {\bfseries\boldmath $0.697$} & $6.32$ & $0.875$ & {\bfseries\boldmath $0.578$} \\
Smallwood int. & -- & $2.64$ & {\bfseries\boldmath $1.56$} & -- & $1.49$ & {\bfseries\boldmath $0.605$} & $2.14$ & $0.849$ & {\bfseries\boldmath $0.456$} & $2.61$ & $0.775$ & {\bfseries\boldmath $0.651$} \\
Smallwood prev. & -- & $2.43$ & {\bfseries\boldmath $1.58$} & -- & $1.42$ & {\bfseries\boldmath $0.617$} & $2.86$ & $0.935$ & {\bfseries\boldmath $0.588$} & $2.39$ & $0.624$ & {\bfseries\boldmath $0.423$} \\
\bottomrule
\end{tabular}
\end{adjustbox}
\caption{\textbf{Reconstruction of the nearest-neighbour degree and strength.} Standardised root mean square error $\mathrm{RMSE}_z=\big[\frac{1}{N}\sum_i (x_i^*-\langle x_i\rangle)^2/\sigma_{x_i}^2\big]^{1/2}$ of $k^{nn}$ and $s^{nn}$ on the two layers, with ensemble averages and standard deviations estimated from sampled networks (Supplementary Note~S3). The smallest error is in bold. BiWCM networks are complete with probability one, so $k^{nn}_i$ is the same in every sample and its $\mathrm{RMSE}_z$ is undefined (--). $^{a}$~Intensity matrix; the prevalence matrix in the source data is identical to it and is not included.}
\label{tab:reconstruction_errors}
\end{table}

%% file: SI_tabs/table_projection_characteristics_v3.tex
\begin{table}[p]
\centering
\scriptsize
\setlength{\tabcolsep}{2.4pt}
\renewcommand{\arraystretch}{1.02}
\begin{adjustbox}{max width=\textwidth}
\begin{tabular}{lccrrrrrrrrr}
\toprule
& & & \multicolumn{3}{c}{Validated fraction} & \multicolumn{3}{c}{F1 overlap} & \multicolumn{3}{c}{Spearman $\rho_s$ of p-values} \\
\cmidrule(lr){4-6}\cmidrule(lr){7-9}\cmidrule(lr){10-12}
Network & Thr. & $S$ (W/P/E) & BiWCM & P-BiECM & BiECM & W--P & W--E & P--E & W--P & W--E & P--E \\
\midrule
Trade 2012 & FDR & $10^3$/$10^3$/$10^3$ & $0.867$ & $0.845$ & $0.774$ & $0.987$ & $0.943$ & $0.956$ & $0.894$ & $0.771$ & $0.875$ \\
Trade 2014 & FDR & $10^3$/$10^3$/$10^3$ & $0.821$ & $0.788$ & $0.704$ & $0.979$ & $0.922$ & $0.942$ & $0.891$ & $0.78$ & $0.889$ \\
Trade 2016 & FDR & $10^3$/$10^3$/$10^3$ & $0.828$ & $0.801$ & $0.717$ & $0.983$ & $0.928$ & $0.944$ & $0.902$ & $0.782$ & $0.883$ \\
Trade 2018 & FDR & $10^3$/$10^3$/$10^3$ & $0.826$ & $0.803$ & $0.723$ & $0.985$ & $0.933$ & $0.947$ & $0.912$ & $0.793$ & $0.885$ \\
Trade 2020 & FDR & $10^3$/$10^3$/$10^3$ & $0.805$ & $0.777$ & $0.687$ & $0.982$ & $0.92$ & $0.938$ & $0.914$ & $0.794$ & $0.886$ \\
\addlinespace[2pt]
Reddit 2013 & FDR & $10^3$/$10^3$/$10^3$ & $0.893$ & $0.891$ & $0.542$ & $0.999$ & $0.755$ & $0.756$ & $0.948$ & $0.551$ & $0.573$ \\
Reddit 2014 & FDR & $10^3$/$10^3$/$10^3$ & $0.865$ & $0.857$ & $0.416$ & $0.995$ & $0.649$ & $0.653$ & $0.881$ & $0.587$ & $0.645$ \\
Reddit 2015 & FDR & $10^3$/$10^3$/$10^3$ & $0.926$ & $0.915$ & $0.512$ & $0.994$ & $0.711$ & $0.715$ & $0.782$ & $0.442$ & $0.527$ \\
Reddit 2016 & FDR & $10^3$/$10^3$/$10^3$ & $0.919$ & $0.911$ & $0.597$ & $0.996$ & $0.787$ & $0.79$ & $0.841$ & $0.481$ & $0.551$ \\
Reddit 2017 & FDR & $10^3$/$10^3$/$10^3$ & $0.917$ & $0.905$ & $0.613$ & $0.993$ & $0.801$ & $0.807$ & $0.81$ & $0.496$ & $0.599$ \\
\addlinespace[2pt]
Lexical 2015 & FDR & $10^3$/$10^3$/$10^3$ & $0.826$ & $0.749$ & $0.696$ & $0.951$ & $0.901$ & $0.933$ & $0.858$ & $0.763$ & $0.867$ \\
Lexical 2018 & FDR & $10^3$/$10^3$/$10^3$ & $0.818$ & $0.744$ & $0.666$ & $0.953$ & $0.889$ & $0.908$ & $0.852$ & $0.713$ & $0.742$ \\
Lexical 2020 & FDR & $10^3$/$10^3$/$10^3$ & $0.805$ & $0.738$ & $0.746$ & $0.957$ & $0.949$ & $0.956$ & $0.867$ & $0.807$ & $0.863$ \\
Lexical 2021 & FDR & $10^3$/$10^3$/$10^3$ & $0.867$ & $0.817$ & $0.782$ & $0.97$ & $0.943$ & $0.949$ & $0.819$ & $0.715$ & $0.723$ \\
Lexical 2023 & FDR & $10^3$/$10^3$/$10^3$ & $0.883$ & $0.847$ & $0.857$ & $0.979$ & $0.969$ & $0.965$ & $0.836$ & $0.761$ & $0.775$ \\
\addlinespace[2pt]
Twitter Q22 & FDR & E/$3{\cdot}10^4$/$3{\cdot}10^4$ & $0.0748$ & $0.0321$ & $7.74\times10^{-3}$ & $0.601$ & $0.187$ & $0.388$ & $0.973$ & $0.978$ & $0.994$ \\
Twitter election & FDR & E/$10^5$/$10^5$ & $0.0295$ & $8.49\times10^{-3}$ & $4.45\times10^{-4}$ & $0.447$ & $0.0296$ & $0.0892$ & $0.965$ & $0.971$ & $0.988$ \\
Twitter crisis & FDR & E/$10^5$/$10^5$ & $0.0149$ & $3.84\times10^{-3}$ & $2.82\times10^{-4}$ & $0.409$ & $0.037$ & $0.0834$ & $0.97$ & $0.981$ & $0.989$ \\
\addlinespace[2pt]
Tech CIT 2010 & FDR & $10^3$/$10^3$/$10^3$ & $0.951$ & $0.894$ & $0.805$ & $0.969$ & $0.917$ & $0.947$ & $0.584$ & $0.512$ & $0.861$ \\
Tech CIT 2012 & FDR & $10^3$/$10^3$/$10^3$ & $0.947$ & $0.897$ & $0.822$ & $0.973$ & $0.93$ & $0.956$ & $0.613$ & $0.533$ & $0.86$ \\
Tech CIT 2014 & FDR & $10^3$/$10^3$/$10^3$ & $0.966$ & $0.925$ & $0.827$ & $0.978$ & $0.922$ & $0.944$ & $0.526$ & $0.438$ & $0.839$ \\
Tech CIT 2016 & FDR & $10^3$/$10^3$/$10^3$ & $0.95$ & $0.898$ & $0.809$ & $0.972$ & $0.92$ & $0.947$ & $0.61$ & $0.524$ & $0.846$ \\
Tech CIT 2018 & FDR & $10^3$/$10^3$/$10^3$ & $0.945$ & $0.873$ & $0.733$ & $0.96$ & $0.874$ & $0.913$ & $0.554$ & $0.479$ & $0.872$ \\
\addlinespace[2pt]
Tech CPC 2010 & FDR & $10^3$/$10^3$/$10^3$ & $0.975$ & $0.95$ & $0.744$ & $0.987$ & $0.865$ & $0.876$ & $0.5$ & $0.275$ & $0.69$ \\
Tech CPC 2012 & FDR & $10^3$/$10^3$/$10^3$ & $0.954$ & $0.92$ & $0.731$ & $0.982$ & $0.865$ & $0.882$ & $0.622$ & $0.396$ & $0.686$ \\
Tech CPC 2014 & FDR & $10^3$/$10^3$/$10^3$ & $0.954$ & $0.932$ & $0.775$ & $0.989$ & $0.897$ & $0.908$ & $0.625$ & $0.367$ & $0.665$ \\
Tech CPC 2016 & FDR & $10^3$/$10^3$/$10^3$ & $0.978$ & $0.954$ & $0.768$ & $0.988$ & $0.88$ & $0.891$ & $0.547$ & $0.282$ & $0.61$ \\
Tech CPC 2018 & FDR & $10^3$/$10^3$/$10^3$ & $0.841$ & $0.588$ & $0.205$ & $0.824$ & $0.392$ & $0.516$ & $0.699$ & $0.602$ & $0.908$ \\
\addlinespace[2pt]
Beehler & 0.05 & $10^3$/$10^3$/$10^3$ & $0.273$ & $0.0364$ & 0 & $0.235$ & 0 & 0 & $0.992$ & $0.971$ & $0.978$ \\
Seed disp. Beehler & 0.05 & $10^3$/$10^3$/$10^3$ & $0.194$ & $0.0833$ & 0 & $0.6$ & 0 & 0 & $0.979$ & $0.96$ & $0.974$ \\
Seed disp. Snow & 0.05 & $10^3$/$10^3$/$10^3$ & $0.287$ & $0.0936$ & $0.0468$ & $0.492$ & $0.246$ & $0.5$ & $0.954$ & $0.948$ & $0.979$ \\
Plant--pollinator BH & 0.05 & $10^3$/$10^3$/$10^3$ & $0.376$ & $0.189$ & $5.82\times10^{-4}$ & $0.671$ & $2.06\times10^{-3}$ & $2.04\times10^{-3}$ & $0.989$ & $0.974$ & $0.975$ \\
Plants--ants & 0.05 & $10^6$/$10^6$/$10^6$ & $0.454$ & $0.102$ & $1.22\times10^{-3}$ & $0.368$ & $5.36\times10^{-3}$ & $0.0235$ & $0.991$ & $0.934$ & $0.952$ \\
Aishihik int. & 0.05 & $10^3$/$10^3$/$10^3$ & $0.19$ & $0.143$ & $0.0476$ & $0.857$ & $0.4$ & $0.5$ & $0.957$ & $0.895$ & $0.849$ \\
Aishihik prev. & 0.05 & $10^3$/$10^3$/$10^3$ & 0 & 0 & 0 & 0 & 0 & 0 & $0.965$ & $0.835$ & $0.829$ \\
Canada & 0.05 & $10^3$/$10^3$/$10^3$ & $0.0222$ & 0 & 0 & 0 & 0 & 0 & $0.975$ & $0.53$ & $0.501$ \\
Cold Lake int. & 0.05 & $10^3$/$10^3$/$10^3$ & $0.0222$ & 0 & 0 & 0 & 0 & 0 & $0.974$ & $0.581$ & $0.549$ \\
Cold Lake prev. & 0.05 & $10^3$/$10^3$/$10^3$ & $0.0667$ & 0 & $0.0222$ & 0 & $0.5$ & 0 & $0.981$ & $0.844$ & $0.822$ \\
Huron prev. & 0.05 & $10^3$/$10^3$/$10^3$ & $0.301$ & $0.116$ & $0.0133$ & $0.555$ & $0.0602$ & $0.0882$ & $0.984$ & $0.884$ & $0.906$ \\
Lake of the Woods & 0.05 & $10^3$/$10^3$/$10^3$ & $0.333$ & $0.0301$ & $0.0108$ & $0.166$ & $0.05$ & $0.316$ & $0.973$ & $0.807$ & $0.812$ \\
Lake of the Woods prev. & 0.05 & $10^3$/$10^3$/$10^3$ & $0.305$ & $0.0366$ & $8.60\times10^{-3}$ & $0.214$ & $0.0411$ & $0.19$ & $0.974$ & $0.917$ & $0.939$ \\
McGregor$^{a}$ & 0.05 & $10^3$/$10^3$/$10^3$ & $0.132$ & $0.0659$ & 0 & $0.667$ & 0 & 0 & $0.906$ & $0.933$ & $0.93$ \\
Parsnip int. & 0.05 & $10^3$/$10^3$/$10^3$ & $0.14$ & $0.0441$ & $0.0294$ & $0.48$ & $0.348$ & $0.6$ & $0.944$ & $0.931$ & $0.971$ \\
Parsnip prev. & 0.05 & $10^3$/$10^3$/$10^3$ & $0.191$ & $0.0662$ & $0.0221$ & $0.514$ & $0.207$ & $0.5$ & $0.972$ & $0.953$ & $0.968$ \\
Smallwood int. & 0.05 & $10^3$/$10^3$/$10^3$ & 0 & 0 & 0 & 0 & 0 & 0 & $0.934$ & $0.934$ & $0.996$ \\
Smallwood prev. & 0.05 & $10^3$/$10^3$/$10^3$ & 0 & 0 & 0 & 0 & 0 & 0 & $0.93$ & $0.909$ & $0.996$ \\
\bottomrule
\end{tabular}
\end{adjustbox}
\caption{\textbf{Validation of the weighted projections on layer 1.} Fraction of validated pairs for each model, F1 overlap between the sets of validated pairs and Spearman correlation between the p-values of all pairs (W = BiWCM, P = P-BiECM, E = BiECM). Column $S$: number of networks sampled from each fitted ensemble to estimate the p-values (W/P/E); E denotes the analytic p-values of the BiWCM on the Twitter matrices (Edgeworth expansion of the exact cumulants, Section~\ref{sec:reconstruction}). Column Thr.: pairs are validated with the Benjamini--Hochberg threshold at $\alpha_{\mathrm{sig}}=0.05$ (FDR) when it exists for all three models, otherwise with the common threshold $\widehat p_{ij}<0.05$ (0.05). All thresholds correspond to at least ten expected exceedances at the given $S$, except for the BiECM on Twitter election and Twitter crisis, where the Benjamini--Hochberg threshold ($2.2\times10^{-5}$ and $1.4\times10^{-5}$) corresponds to about two exceedances at $S=10^5$, and the validated sets of that model may still shrink with more samples. The p-value correlations do not depend on the threshold. $^{a}$~Intensity matrix; the prevalence matrix in the source data is identical to it and is not included.}
\label{tab:projection_characteristics}
\end{table}

%% file: SI_tabs/table_benchmark_v2.tex
\begin{table}[t]
\centering
\scriptsize
\setlength{\tabcolsep}{4pt}
\begin{tabular}{llrrrrr}
\toprule
Model & Solver & Converged & \multicolumn{2}{c}{Time (s)} & \multicolumn{2}{c}{Iterations} \\
\cmidrule(lr){4-5}\cmidrule(lr){6-7}
 & & (\%) & median & max & median & max \\
\midrule
BiCM & Fixed point & 100 & 0.07 & 20.6 & 949 & 1316 \\
 & Quasi-Newton & 100 & $8.4\times10^{-3}$ & 22.7 & 25 & 57 \\
 & Coordinate & 100 & 0.11 & 229 & 6 & 14 \\
\addlinespace[2pt]
BiWCM & Fixed point & 98 & 0.01 & 98.5 & 41 & 599 \\
 & Quasi-Newton & 87 & $6.1\times10^{-3}$ & 92.2 & 43 & 90 \\
 & Coordinate & 100 & 0.28 & 8259 & 24 & 87 \\
\addlinespace[2pt]
P-BiECM & Fixed point & 96 & 0.03 & 274 & 189 & 1797 \\
 & Quasi-Newton & 73 & 0.02 & 480 & 50 & 157 \\
 & Coordinate & 100 & 0.47 & 55725 & 24 & 225 \\
\addlinespace[2pt]
BiECM & Fixed point & 93 & 0.10 & 47.9 & 594 & 1790 \\
 & Quasi-Newton & 100 & 0.31 & 13292 & 448 & 3746 \\
 & Coordinate & 100 & 1.1 & 41155 & 26 & 188 \\
\addlinespace[2pt]
qBiECM & Fixed point & 89 & 0.06 & 200 & 1084 & 3365 \\
 & Quasi-Newton & 82 & 0.02 & 2779 & 102 & 1393 \\
 & Coordinate & 100 & 0.34 & 247395 & 22 & 1511 \\
\bottomrule
\end{tabular}
\caption{\textbf{Convergence and computational cost of the solvers.} Each model is fitted with the three solvers on the 45 empirical networks of Fig.~\ref{fig:benchmark} (Twitter crisis excluded). A run is converged when the maximum relative error on the constraints falls below $10^{-3}$ within 5000 iterations. Times and iterations refer to converged runs (single CPU run per network and solver).}
\label{tab:benchmark}
\end{table}